\documentclass[useAMS,usenatbib,aas_macros]{mn2e}

\usepackage{epsfig}
\usepackage{lscape,graphicx,colortbl}
\usepackage{amssymb}
\usepackage{natbib}
\usepackage{times}
\usepackage{aas_macros}

\newcommand{\msun}{M_\odot}
\newcommand{\ifm}[1]{\relax\ifmmode#1\else$\mathsurround=0pt#1$\fi}
\newcommand{\kms}{\ifmmode\,{\rm km}\,{\rm s}^{-1}\else km$\,$s$^{-1}$\fi}

\newcommand{\hmsun}{\,\ifm{h^{-1}}{M_{\odot}}}
\def\omm{\Omega_{\rm m}}
\def\oml{\Omega_{\Lambda}}

\newcommand{\be}{\begin{equation}}
\newcommand{\ee}{\end{equation}}
\newcommand{\bea}{\begin{eqnarray}}
\newcommand{\eea}{\end{eqnarray}}
\newcommand{\z}{\emph{z}}

\def\m{{\bf m}}
\def\A{{\bf A}}
\def\B{{\bf B}}
\def\ms{m_{\rm star}}
\def\mc{m_{\rm cold}}
\def\mh{m_{\rm hot}}

\def\fs{f_{\rm s}}
\def\fe{f_{\rm e}}
\def\fr{f_{\rm re}}
\def\fsdb07{f_{\rm s,D}}
\def\fc{f_{\rm c}}

\def\ffd{f_{\rm d}}

\newcommand{\fof}{{\scshape fof~}}

\def\mfal{M_{\rm infall}}  
\def\zfal{\z_{\rm infall}}  
\def\mf{M_{\rm 200}} 
\def\msh{m_{\rm star}-M_{\rm infall}}  

\begin{document}

\title[Linking haloes to galaxies]
      {Linking haloes to galaxies: how many halo properties are needed?}

\author[E. Neistein et al.]
{Eyal Neistein$^{1,2}$\thanks{E-mail:$\;$eyal@mpe.mpg.de}, 
Simone M. Weinmann$^{1,3}$, 
Cheng Li$^{4,1}$,
Michael Boylan-Kolchin$^{1,5}$\\
$^{1}$Max-Planck-Institute for Astrophysics, Karl-Schwarzschild-Str. 1,
85748 Garching, Germany \\ 
$^{2}$ Max-Planck-Institute for Extraterrestrial Physics,
Giessenbachstrasse 1, 85748 Garching, Germany\\
$^{3}$ Leiden Observatory, Leiden University, P.O. Box 9513, 2300 RA Leiden,
The Netherlands \\ 
$^{4}$ Max-Planck-Institute Partner Group, Key Laboratory for Research in
Galaxies and Cosmology, \\ \quad \quad Shanghai Astronomical Observatory, Chinese
Academy of Sciences, Nandan Road 80, Shanghai 200030, China \\
$^{5}$Center for Galaxy Evolution, 4129 Reines
Hall, University of California, Irvine, CA 92697, USA}


\date{}
\pagerange{\pageref{firstpage}--\pageref{lastpage}} \pubyear{2010}
\maketitle

\label{firstpage}


\begin{abstract}
Recent studies emphasize that an empirical relation between the
stellar mass of galaxies and the mass of their host dark matter
subhaloes can predict the clustering of galaxies and its evolution with 
cosmic time.
In this paper we study the various assumptions made by this methodology
using a semi-analytical model (SAM). To this end, we randomly swap between
the locations of model galaxies within a narrow range of subhalo mass ($\mfal$).
We find that shuffled samples of galaxies have different auto-correlation
functions in comparison with the original model galaxies. This difference is
significant even if central and satellite galaxies are allowed to follow a different 
relation between $\mfal$ and stellar mass, and can reach a factor of $\sim2$ for 
massive galaxies at redshift zero.
We analyze three features within SAMs that contribute to this effect:
a) The relation between stellar mass and subhalo mass evolves with
redshift for central galaxies, affecting satellite galaxies at
the time of infall.
b) In addition, the stellar mass of galaxies falling into groups and clusters at 
high redshift is different from the mass of central galaxies at the same time.
c) The stellar mass growth for satellite galaxies after infall can be significant and
depends on the infall redshift and the group mass.
All of the above ingredients modify the stellar mass of satellite galaxies in
a way that is more complicated than a dependence on the subhalo mass only.
By using two different SAMs, we show that the above is true for differing models of 
galaxy evolution, and that the effect is sensitive to the treatment of dynamical friction 
and stripping of gas in satellite galaxies.
We find that by using the \fof group mass at redshift zero in addition to
$\mfal$, an empirical model is able to accurately reproduce the clustering
properties of galaxies. On the other hand, using the infall redshift as a second parameter does not
yield as good results because it is less correlated with stellar mass.
Our analysis indicates that environmental processes that affect galaxy evolution are
important for properly modeling the clustering and abundance of galaxies.
\end{abstract}




\section{Introduction}
\label{sec:intro}

The current paradigm for the formation of structure in the Universe predicts that galaxies
form and evolve inside of dark matter haloes.  This working assumption makes the
relation between the properties of galaxies and their host haloes a fundamental
probe for various aspects of galaxy formation and cosmology. Galaxy properties
such as stellar mass, star formation rate, color, and clustering are all
intimately linked to the host halo mass. Consequently, there have been many
recent attempts to quantify the relation between the mass of galaxies and their
host halo mass (hereafter the `mass relation').

Observational constraints on the mass relation were derived by weak lensing
\citep[e.g.][]{Mandelbaum06}, satellite dynamics \citep[e.g.][]{Conroy07} and
large group catalogs of galaxies \citep{Yang09}. All these are able to provide
constraints on the stellar-halo mass relation for $10^{12}-10^{14} \msun$
haloes. At lower halo masses there are
currently no useful direct observational constraints. Nonetheless, the
abundance of low mass haloes is important for constraining the nature of
dark matter \citep[see, e.g.][]{Bode01,Maccio10}.

From the theoretical perspective, there are two main approaches for studying the
relationship between stellar and halo masses. A straightforward way is to use
existing models and extract the mass relation as a secondary result.  This
approach was demonstrated using hydrodynamical simulations
\citep[e.g.][]{Sawala10}, and semi-analytical models
\citep[][]{Wang06,Moster10}. These two methodologies result in very different
mass relations: hydrodynamical simulations predict higher stellar mass for
a given halo mass, especially for low mass haloes. It is not clear whether the
origin of this discrepancy is the different set of assumptions adopted by each
model, the different observational data which are used to constrain the models,
or some specific physical ingredients.

A different approach is to use the stellar-halo mass relation as the main assumption,
and to verify it against a variety of observations. This approach is more useful for
a systematic analysis of the mass relation, and for carefully testing the necessary
ingredients needed to explain it. The halo occupation
distribution (HOD) was historically the first approach to follow this
line of thought \citep[see][and references therein]{Berlind02,Tinker05,Zehavi05}.
It assigns per each halo a specific \emph{number} of galaxies of a given
type. By tuning the parameters of the number and location distributions, these
models are able to reproduce the abundance and clustering properties of galaxies.
Thus, the relation between haloes and galaxies can be constrained using a relatively simple
set of assumptions, and the halo mass is a sufficient parameter for fixing
the properties of its galaxies.

Various studies have shown that the HOD approach can be simplified even more
using information on the substructure within haloes, i.e., the \emph{subhalo}
mass \citep{Kravtsov04,Vale04,Conroy06,Shankar06,Guo10a,Moster10}. In this
approach, each subhalo can host only one galaxy. The galaxy mass is linked to
the subhalo mass at the last time the subhalo was central within its \fof group
(hereafter $\mfal$). The specific relation between stellar mass and $\mfal$ is
then fixed by matching the abundance of galaxies and $\mfal$.  This method is
therefore termed `abundance matching' (ABM; models based on the ABM approach
will be termed ABMs in what follows).  Using subhalo samples from large
cosmological $N$-body simulations, ABMs are able to reproduce the
auto-correlation function of galaxies surprisingly well.
The main difference between the HOD and the ABM approach is thus that
HOD uses the \fof mass for determining stellar
masses, while the ABM approach uses the subhalo mass, which is limited to contributions from
smaller scales of dark matter.

By fixing the stellar mass according to $\mfal$ only, ABMs assume various
simplifications to the statistics of galaxy formation.  First, they neglect the
redshift where the galaxy became a satellite.  This redshift should affect the
stellar mass because the mass relation evolves with redshift.
Second, the stellar mass gained by satellite galaxies
after infall is assumed to be independent of clustering.  Third, ABMs do not
allow a galaxy's large-scale environment to modify its stellar mass
\citep{Croton07}.  However, it seems that these simplifications do not force the
models to violate the clustering properties of galaxies. It is not clear if this
is due to the low sensitivity of the auto-correlation function to these
assumptions, or because multiple effects conspire to cancel each other and leave
the auto-correlation function unchanged.

In this paper we will test the ABM assumptions using detailed
semi-analytical models (hereafter SAMs).
We will test whether the stellar mass function and clustering 
of galaxies in the SAMs depend only on $\mfal$.
It will be shown that SAMs do not follow the above assumptions made
by ABM models. The inclusion of more detailed
physics, and the fact that SAMs follow the evolution of each individual galaxy, affects the
auto correlation functions of the model galaxies.
The success of ABMs thus poses important questions on
the processes which govern galaxy formation. Is it possible that the simple
ABM approach mimic the observed clustering better than the more sophisticated
models? Or is their success purely a coincidence?

In order to compare ABMs to SAMs we will use the technique of shuffling
between the location of model galaxies. This method was introduced by \citet{Croton07},
as a tool to quantify the effect of large-scale environment on the stellar mass of galaxies.
These authors found that environment of the host \fof halo
can contribute up to $\sim10$ per cent
variation in the correlation function of galaxies.
Here we adopt a similar technique in order to test the role of $\mfal$ in shaping
the stellar mass of galaxies and their correlation function.
The effect we find is an order of magnitude larger than the one
found by \citet{Croton07} at least for high mass galaxies.
This is because Croton et al. (2007) test the HOD model, and not ABM,
which means that they shuffle galaxies between \fof groups and not between subhaloes.

One of the targets of this paper is to suggest possible ingredients for more
detailed ABM studies. We would like to find which ingredients might contribute
to the clustering of galaxies, independent of the specific model chosen
here. This will allow future ABM studies to fully explore the parameter
space. For this reason, we use models that span a large range of possible
physical recipes. The two basic SAMs we use in the work are taken from \citet{Neistein10}, 
which discusses six different specific models spanning a large range of galaxy formation
scenarios.

This paper is organized as follows. In section \ref{sec:method} we
describe the two different methodologies used in this work, namely
ABMs and SAMs. The shuffling test, which reveals deviations between
ABMs and SAMs, is discussed in section \ref{sec:shuffle}. Section \ref{sec:differences}
is devoted to a more systematic analysis of the various differences
between the two methodologies. In section \ref{sec:add_param} we then
suggest an extension to ABMs that is able to reproduce the results of SAMs. Lastly,
we summarize the results and discuss them in section
\ref{sec:discuss}.

This paper is based on the cosmological model with
$(\omm,\,\oml,\,h,\,\sigma_8)=(0.25,\,0.75,\,0.73,\,0.9)$, which is
the model adopted by the Millennium simulation used here
\citep{Springel05}. Mass units are $\msun$ unless otherwise noted; Log
designates Log$_{10}$.

\section{Methodologies}
\label{sec:method}

\subsection{Abundance matching}
\label{sec:abm_method}

Abundance matching models (ABMs) link the stellar mass of
a galaxy to its host subhalo mass using a simple and empirical methodology.
The ingredients of the models can be summarized as follows:
\begin{enumerate}
  \item Subhalo selection and mass definition: here we choose which subhaloes
  are allowed to host galaxies, and how their mass is defined. As a result,
  the abundance of subhaloes and their locations are set.
  \item Assume a one-to-one relation between the subhalo mass
  and the stellar mass of its galaxy: $\msh$. More complex models
  allow scatter, or use different relations for satellite and central
  galaxies.
  \item Solve for the $\msh$ relation, in order to reproduce the
  observed stellar mass function.
\end{enumerate}
For simplicity, we discuss ABMs which are aimed at producing a
population of galaxies at $\z=0$ only. Such models are usually
tested against the auto-correlation function of galaxies
at $\z=0$ (see \S\ref{sec:computing_CF} below). For a detailed review of ABMs and the
various uncertainties involved see \citet{Behroozi10}.
In this section we further explain the various ingredients of ABMs
and specify the actual model assumed here.

We first define the subhalo mass, $\mfal$, which will be used later
in order to fix the stellar mass of galaxies:
\begin{equation}
\label{eq:minfall}
\mfal = \left\{ \begin{array}{ll}
M_h & \textrm{if central within its \fof group}\\ \;\;\;\
& \;\;\;\;\;\;\;\;\;\;\;\;\;\;\;\;\;\;\;\;\;\;\;\;\;\;\;\;\;\;\;\;\;\;\;  \\
M_{h,p}(\zfal) & \textrm{otherwise}
\end{array} \right.
\end{equation}
Here $\zfal$ is the lowest redshift at which the main progenitor\footnote{
Main-progenitor histories are derived by
following back in time the most massive progenitor in each merger event.}
of the subhalo $M_h$
was the central of its \fof group, and $M_{h,p}$ is the main progenitor mass at
this redshift. We select all the subhaloes from the Millennium simulation
\citep[][see section \ref{sec:trees}]{Springel05} at $\z=0$.
Unlike many ABMs, we also consider subhaloes that are not identified at $\z=0$ if their galaxies
survive within the SAM and did not have enough time to merge with the central
galaxy. Even though these subhaloes are only identified at high redshift, their $\mfal$ can
still be computed according to Eq.~\ref{eq:minfall}.

The basic assumption of ABMs \citep[e.g.][]{Conroy06} is listed in item
(ii) above: \emph{the
stellar mass of a galaxy depends solely on $\mfal$}. Most ABMs assign
a unique
$\ms$ value per each subhalo mass. However, several studies
allow a scatter in the mass relation, and \citet{Wang06}
use different relations for central galaxies versus
satellite galaxies (the mass relations are adopted from the results
of a SAM in this last case). In this work we mainly
discuss the most detailed ABM approach, where the $\msh$ relation
includes scatter, and might be different for satellite galaxies.
We will show that even when using this extended approach, the SAM
behaviour is different from ABMs.

Once the assumption on the nature of the $\msh$ relation
has been made, the last step in constructing the model is to
find a relation $\msh$ that will reproduce the observed abundance of $\ms$.
If all the subhaloes from item (i) are populated with
galaxies according to this relation, the resulting set of galaxies
would have the observed stellar mass function by construction. The $\msh$ relation
can be thus obtained if one assumes that it follows some general functional
shape, where the free parameters are constrained by matching the $\ms$ abundance.
A different solution can be obtained by solving for the numerical values for
$\msh$ in each mass bin separately. In
this work we do not assume any functional shape a-priori, but rather
adopt the same $\msh$ relation as in the SAMs.

Although ABMs seem to adopt a rather simplified set of assumptions, it turns
out that the clustering properties of galaxies fit the
observational data well. This is encouraging, as clustering is the
main test of the model (stellar mass functions are reproduced by
construction).
In addition, it was shown by \citet{Wang06} that the results of a specific
SAM agrees quite well with this approach. All of these tests indicate
that ABMs provide a reasonable solution to the relation between halo
mass and stellar mass. It is not clear, however, whether this
relation is the \emph{only} possible solution.


\subsection{The Semi-analytical models}
\label{sec:sam}

In this section we briefly describe the SAM formalism being used in this work for
modeling the evolution of galaxies. For more details the reader is referred
to \citet{Neistein10} (hereafter NW10). A version of the code used in this work
is available for public usage through the internet
(at \texttt{http://www.mpa-garching.mpg.de/galform/sesam}).

\subsubsection{Merger trees}
\label{sec:trees}

We use merger trees extracted from the Millennium
$N$-body simulation \citep{Springel05}. This simulation was run
using the cosmological parameters
$(\omm,\,\oml,\,h,\,\sigma_8)=(0.25,\,0.75,\,0.73,\,0.9)$, with a
particle mass of $8.6\times10^8\,\hmsun$ and a box size of 500
$h^{-1}$Mpc.
The merger trees used here are based on \emph{subhaloes} identified using the
\textsc{subfind} algorithm \citep{Springel01}. They are defined as the bound
density peaks inside \fof groups \citep{Davis85}.
More details on the simulation and the subhalo
merger-trees can be found in \citet{Springel05} and
\citet{Croton06}. The mass of each subhalo (referred to as $M_h$ in what follows)
is determined according to the number of particles it contains.
Within each \fof group the most massive subhalo  is termed
the central subhalo of its group.

\subsubsection{Quiescent evolution}

Each galaxy is modeled by a 4-component vector,
\begin{eqnarray}
\m = \left( \begin{array}{c} \ms \\ \mc \\ \mh \\ m_{\rm eject} \end{array} \right)
\, ,
\label{eq:m_vec}
\end{eqnarray}
where $\ms$ is the mass of stars, $\mc$ is the mass of cold gas,
$\mh$ is the mass of hot gas distributed within the host subhalo, and $m_{\rm eject}$
is the mass of gas which is located out of the subhalo and is not able to cool directly
into the cold phase. We use the term `quiescent evolution' to mark all the evolutionary
processes of a galaxy except those related to mergers.

It can be shown (see NW10) that most of
the quiescent processes included in SAMs can be
written in a compact form by using linear differential equations.
We therefore adopt the following model for the quiescent evolution:
\begin{equation}
\dot{\m} = \A\m + \B\dot{M}_h \, ,
\label{eq:m_evolve}
\end{equation}
where $\dot{M}_h$ is the growth rate of the subhalo mass due to smooth accretion
(i.e., mass which does not come within other subhaloes), and
\begin{eqnarray}
\nonumber
\A = \left( \begin{array}{cccc}
0 & (1-R)\fs & 0 & 0 \\
0 & -(1-R)\fs -\ffd \fs & \fc & 0 \\
0 & \ffd \fs - \fe \fs & -\fc & \fr \\
0 & \fe \fs & 0 & -\fr
\end{array} \right)
\end{eqnarray}
\begin{eqnarray}
\B = \left( \begin{array}{c}
0  \\
0  \\
0.17 \\
0
\end{array} \right)  \,\, .
\label{eq:AB_defs}
\end{eqnarray}

In short, $\fs,\,\fc,\,\ffd,\,\fe,\,\fr$ are all functions of subhalo
mass and redshift, and correspond to the
efficiencies of star-formation, cooling, feedback, ejection, and
reincorporation respectively. $R$ is the constant recycling
factor, which has the values of $0.43$ or $0.5$ in all the models used in this
work.

\subsubsection{Satellite galaxies: dynamical friction, stripping, and bursts}
\label{sec:formalism_mergers}

Satellite galaxies are defined as all galaxies inside
a \fof group except the main galaxy inside
the central (most massive) subhalo. Once the subhalo corresponding
to a given galaxy cannot be resolved anymore, it is considered
as having merged into the most massive subhalo which has the same descendant (the
`target' subhalo). Due to the effect of dynamical friction,
the galaxy is then assumed to spiral towards the center of the target subhalo,
and merge with the central galaxy in that subhalo after a
(potentially significant) delay time. We thus divide galaxies into three different types:
\begin{equation}
\label{eq:gal_type}
\textrm{galaxy type} = \left\{ \begin{array}{ll}
0 & \textrm{if central within its \fof group}\\
1 & \textrm{if central within its subhalo (`satellite')}\\
2 & \textrm{otherwise (`satellite')}
\end{array} \right.
\end{equation}
Galaxies of type 0 and 1 can serve as merging targets for type 2 galaxies. A type 2 galaxy
is not following the evolution of its original host subhalo,
as its subhalo is not identified anymore. Consequently, type 2 galaxies do not
have a well defined location given by the simulation data. We estimate the location of these
galaxies by using the location of the most bound particle of the last identified subhalo.
This method was used before by e.g. \citet{Croton06,Guo10}.

In our model, the time it takes the galaxy to fall into the central galaxy is determined by
dynamical friction, and is not directly related to the true
evolution of the location of the most bound particle with
respect to its target.
At the last time the dark matter subhalo of a satellite galaxy is resolved
we compute its distance from the
target subhalo ($r_{\rm sat}$), and estimate the dynamical friction
time using the formula of \citet{Binney87},
\begin{equation}
t_{\rm df} = \alpha_{\rm df} \cdot \, \frac{1.17 V_v r_{\rm sat}^2}{G m_{\rm sat}\ln\left(
1+ M_h/m_{\rm sat} \right) } \, .
\label{eq:t_df}
\end{equation}
Here $M_h$ is the mass of the target subhalo and $V_v$ is its virial velocity.
The value of $m_{\rm sat}$ should correspond to the mass of the satellite galaxy
which is affected by the dynamical friction process. We use two options
for $m_{\rm sat}$ as listed below:
\begin{equation}
\label{eq:msat}
m_{\rm sat} = \left\{ \begin{array}{ll}
m_{\rm star}+m_{\rm cold}+M_{h,min} & (a) \\
M_h & (b) \,\,.
\end{array} \right.
\end{equation}
Here $M_{h,min}$ is the minimum
subhalo mass of the Millennium simulation ($1.72\times10^{10}\hmsun$), and
$M_h$ is the last subhalo mass identified, just before the subhalo has merged
into a bigger one.
In addition to the freedom in choosing $m_{\rm sat}$, a free parameter, $\alpha_{\rm df}$,
was added to Eq.~\ref{eq:t_df} in order to easily modify the dynamical friction time.
When a satellite falls into a larger subhalo together with its central
galaxy we update $t_{\rm df}$ and the target subhalo, for both objects according to
the new central galaxy.

While satellite galaxies move within their \fof group, they suffer
from mass loss due to tidal stripping. Only the ejected and hot gas of the
satellite can be stripped in our model. We assume that this stripping has an exponential
dependence on time, using the same time scale for all galaxies.
In order to model this stripping we modify $\A$
by subtracting a constant $\alpha_h$ from two of its elements:
\begin{eqnarray}
\A_{sat}(3,3) &=& -\fc-\alpha_h \\ \nonumber
\A_{sat}(4,4) &=& -\fr-\alpha_h \,.
\label{eq:ABsat_defs}
\end{eqnarray}
This constant suggests an exponential decrease in the
amount of hot and ejected gas. However, the actual
dependence of these components on time is more
complicated due to contributions from other processes, as
seen in Eq.~\ref{eq:AB_defs}.

When satellite galaxies finally merge we assume that a SF burst is
triggered. We follow \citet{Mihos94,Somerville01,Cox08} and
model the amount of stars produced by
\begin{equation}
\Delta \ms = \alpha_b \left( \frac{m_1}{m_2} \right)^{\alpha_c} (m_{1,{\rm cold}}+m_{2,{\rm cold}}) \, .
\label{eq:sf_burst}
\end{equation}
Here $m_i$ are the baryonic masses of the progenitor galaxies (cold
gas plus stars), $m_{i,{\rm cold}}$ is their cold gas mass, and
$\alpha_b$, $\alpha_c$ are constants.

\subsubsection{Correlation functions}
\label{sec:computing_CF}

\begin{figure*}
\centerline{\psfig{file=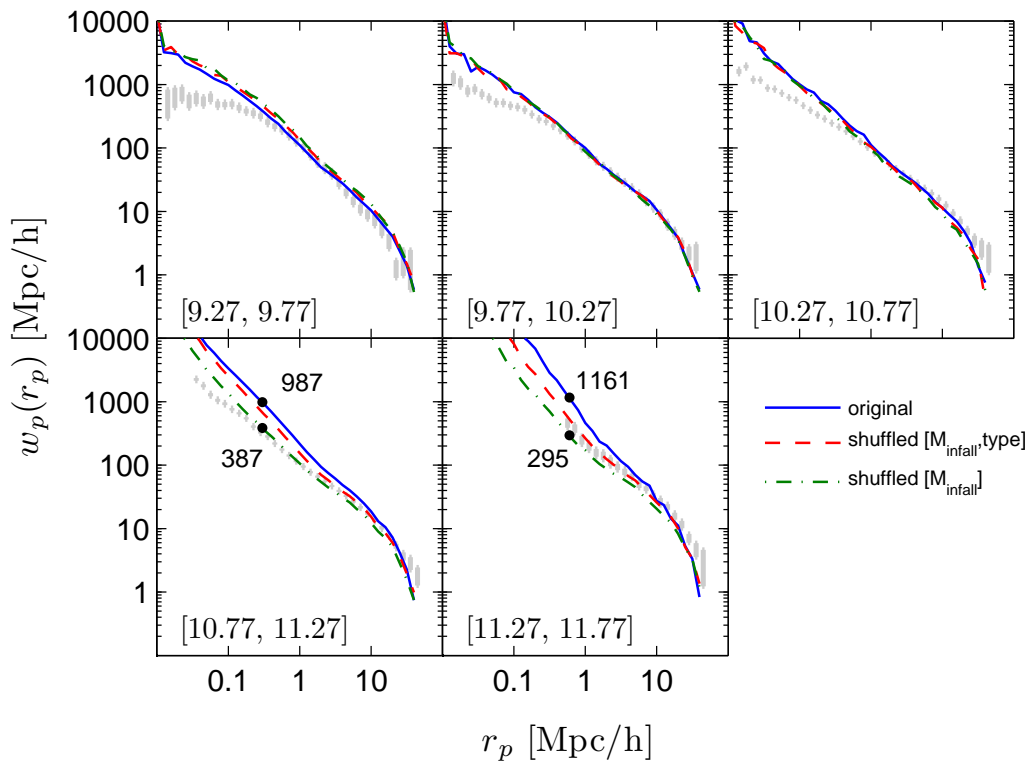,width=150mm,bbllx=30mm,bblly=80mm,bburx=188mm,bbury=200mm,clip=}}
\caption{The projected auto-correlation functions derived for model $A0$.
Each panel corresponds to galaxies with stellar masses as indicated by the range of Log$\msun$.
Solid lines show the results of the original model, dashed lines are
plotted using shuffling within [$\mfal$,type], dashed-dotted lines
represent shuffling within [$\mfal$] only.
See \S\ref{sec:shuffle} for more information about the shuffling procedure. The observational
data are using SDSS DR7 with the same technique as in \citet{Li06}, and are shown as error bars.
In the two most massive bins we add labels for the y-axis values of the corresponding line.
Poisson errors for the model results are smaller than 10 percent for all points, 
and would not be visible in this plot.}
\label{fig:cf_shuffled_A0}
\end{figure*}

In this paper we study the clustering properties of various models using the projected
auto-correlation function, $w_p(r_p)$, termed CF hereafter. Observational data were
obtained from the full SDSS DR7 release, using 1/Vmax weighting, in the same method
as described in \citet{Li06} and presented in \citet{Guo10}. Errors are estimated
from a set of 80 mock SDSS surveys mimicking cosmic variance effects.
The CFs are split into five stellar mass bins, and
are plotted as error-bars in Fig.~\ref{fig:cf_shuffled_A0} (the same data points are used
throughout this work).

In order to compute $w_p(r_p)$ for the models, we use only two coordinates for the
location of model galaxies (i.e. $r_x,r_y$, omitting the $r_z$ dependence).
We then count the number of galaxy pairs, $N_p$, within the same stellar mass bin,
and at a given separation $r_p$ ($r_p=\sqrt{r_x^2+r_y^2}$). The CF is computed from
$N_p$ using:
\begin{equation}
w_p(r_p) = \left[ \frac{L^2}{N^2} \frac{N_p}{V_p} - 1 \right] L \,.
\end{equation}
Here $V_p$ is the 2-dimensional area covered by the bin $r_p$, $N$ is the
total number of galaxies in the sample, and $L$ is the size of the simulation box in $h^{-1}$Mpc.
The resulting $w_p(r_p)$ has units of $h^{-1}$Mpc.
In order to save computational time, we compute CFs only for a random subset of
10 per cent of the galaxies within
the lower three mass bins (we use all galaxies for the two most massive bins).
We checked that there is no difference between the CF computed using the partial
sample and the full sample from the Millennium simulation used here.

\subsubsection{Specific models}
\label{sec:models}

\begin{table}
\caption{The different SAM models used in this paper. The models
are divided into two groups, $A$ \& $B$. Within each group the
specific models are the same except the details given in this table.
Different $m_{\rm sat}$ mass
estimates for computing dynamical friction are given in Eq.~\ref{eq:msat}.
Model $A4$ evolves satellite galaxies
according to \citet{Weinmann10}}
\begin{center}
\begin{tabular}{lccccc}
\hline Name & $\alpha_h^{-1}$ [Gyr] & $\alpha_{\rm df}$ & $m_{\rm sat}$ &  Other  \\
\hline
$A0$ & 4     & 3   & a &  \\
$A1$ & 0     & 3   & a & no stellar growth for satellites \\
$A2$ & 0.01 & 2   & b  & \\
$A4$ & 4 (for type 2)    & 0.3 & b  & \citet{Weinmann10} \\
$\,$ &  $\,$ & $\,$ & $\,$ &  \\
$B0$ & 0.01 & 2 & b & \\
$B1$ & 0 & 2 & b & no stellar growth for satellites \\
$B2$ & 4 & 3 & a  & \\
\hline
\end{tabular}
\end{center}
\label{tab:models}
\end{table}

We choose to examine the results of two models which were originally presented in
NW10. Other models from NW10 have similar results and are not adding a significant
information to this work. We use model II
from NW10, in which SF, feedback, and mergers are treated in a similar way as in
standard SAMs. Three main modifications to usual SAMs are adopted in this model:
i) There is no ejection of gas out of the subhalo ($m_{\rm eject}\equiv0$); ii) the
SF law does not include a threshold in cold gas density; iii) Cooling rates ($\fc$) are tuned
to reproduce a large set of observational data. This model will be
termed model $A0$ in this paper. The second model being used here is the one
based on \citet{DeLucia07} (model 0 in NW10), and is termed model $B0$.

There are a few differences between the results of models $A0$ and $B0$ that are
important for the following sections. The stellar mass functions of these models
are very different (see Fig.~\ref{fig:mass_funs} in the Appendix). Model $A0$
fits the observed
stellar mass functions for $\z<3$, while model $B0$ shows significant
deviations at low and high masses.  We will see below that this difference
affects the relation between the subhalo mass and stellar mass, and the CFs. The
treatment of satellite galaxies is also different between these models. The
dynamical friction and stripping time-scales are longer in model
$A0$. Consequently, satellite galaxies survive longer, and are able to form a
significant amount of stars after joining the group. Longer dynamical friction
time-scales also result in more galaxies of type 2 in model $A0$.

As we will show later in this work, the results of the CFs are sensitive to the treatment
of dynamical friction and gas stripping in satellite galaxies.
In order to further investigate these effects we run a few variations of the
original models. Models $A1$,$B1$ are the same as models `0' except
that satellite galaxies are not allowed to grow in their stellar mass.
We artificially shut down all modes of SF and do not allow merging into
galaxies of type 1 \& 2. These models mimic extremely
fast stripping mechanisms.

In models $A2$,$B2$ we exchange the recipes of dynamical
friction and stripping between models $A0$ \& $B0$, showing the effect of these
ingredients on the results. This will help us disentangle between
the effects of quiescent evolution, and treatment of satellites.

Lastly, in model $A4$, we use a prescription for the evolution of satellite
galaxies as proposed by \citet{Weinmann10}.  In this model gas stripping follows
the stripping of the subhalo mass (as calculated from the $N$-body simulation),
where the first quantity to be stripped is $m_{\rm eject}$. Only when $m_{\rm
  eject}$ reaches zero does stripping of the hot gas start. For type 2 galaxies,
we use exponential stripping of hot gas, using $\alpha_h$ defined in Eq.~\ref{eq:ABsat_defs}
above. Model $A4$ was
developed in order to match in detail the properties of satellite galaxies, and
is thus the most physically motivated model here, in terms of environmental
effects within the halo. When tuning the dynamical friction time scale in Model $A4$ we have
tried to better reproduce the observed CFs (see Fig.\ref{fig:cf_shuffled_A4} in
the Appendix).

Table \ref{tab:models} summarizes the various models used in this work.


\section{The shuffling test}
\label{sec:shuffle}

\begin{figure}
\centerline{ \hbox{ \epsfig{file=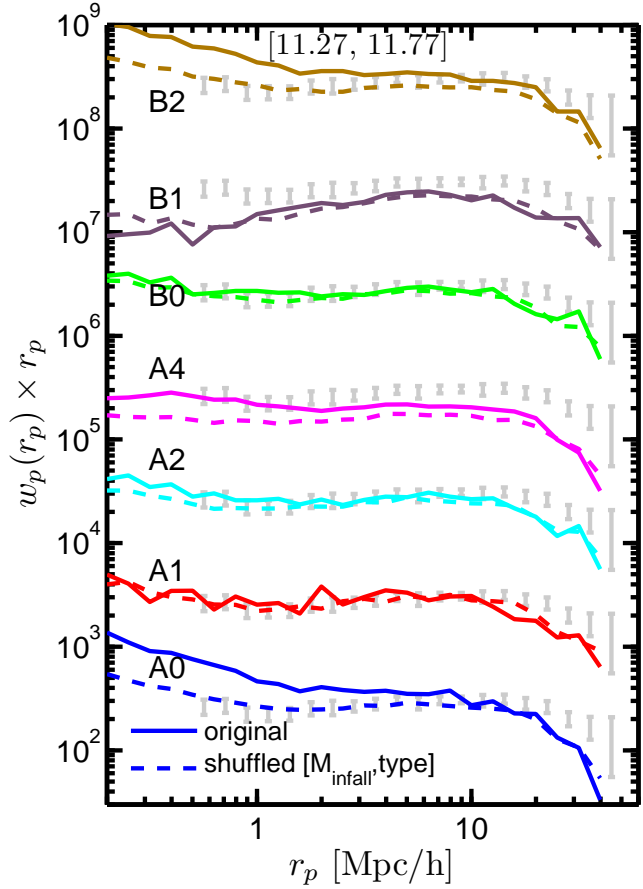,width=9cm} }}
\caption{The projected auto-correlation function for the models listed in
Table \ref{tab:models}. Only galaxies with stellar mass $11.27\leq\ms\leq11.77$ Log$\msun$ are selected
(corresponding to the most massive stellar bin in Fig.~\ref{fig:cf_shuffled_A0}).
\emph{Solid} lines are the original (un-shuffled) CFs. \emph{Dashed} lines show the shuffling
results of each model (shuffling is preformed within galaxies of the same
$\mfal$ and type). The CFs of different models are multiplied by powers of 10 for clarity.}
  \label{fig:massive_cfs}
\end{figure}

We want to check whether SAMs include some clustering information
that is lost by the assumptions of ABM.  A simple test to this
problem is to adopt the same $\msh$ relation as in the
SAM, and use it within an ABM application. Such a test was done by
\citet{Wang06}. 
Here we choose to use the shuffling procedure for testing ABMs against SAMs. In
general, we will randomly re-distribute the location of SAM galaxies with the
same $\mfal$.  In this way we mimic an ABM model with the same subhaloes as used in
the SAM, and with exactly the same distribution of $\ms$ per a given
$\mfal$.
Information about the clustering of galaxies that is not related to $\mfal$ should be 
eliminated by this procedure, and its effect on the CFs can be easily tested
without modeling the specific $\msh$ relation.  Moreover, it will be easy to
see how different modifications of SAMs behave under shuffling, and to add
constraints on the shuffled groups.

We first run the SAMs and construct a catalog of galaxies at $\z=0$. For each
galaxy we save its location at $\z=0$, its type (0/1/2, see
Eq.~\ref{eq:gal_type}), its host $\mfal$, the infall redshift $\zfal$, and the
host \fof group mass at $\z=0$, $\mf$\footnote{$\mf$ is defined as the mass
  within the radius where the halo has an over-density 200 times the critical
  density of the simulation.}.  We then split the population of galaxies into
groups of the same $\mfal$\footnote{We allow $\mfal$ to deviate in 0.1 dex, so
  each group includes a large number of galaxies. We have verified that using
  bin sizes of 0.05-0.2 do not affect the results presented here.}. Within each
group of galaxies we randomly re-distribute the stellar masses at the group
locations.  In order to further explore the clustering properties of SAMs, we
will refer in this work to a few versions of shuffling:
\begin{itemize}
  \item Shuffling within [$\mfal$]
  \item Shuffling within [$\mfal$,type]
  \item Shuffling within [$\mfal$,type,$\zfal$]
  \item Shuffling within [$\mfal$,type,$\mf$]
\end{itemize}
In each case we split the catalog of galaxies into groups where the
values of the variables listed in square brackets are the same
\footnote{The bins used for $\zfal$ are
of size 0.1 in Log$z$, and bins for $\mf$ are of size 0.5 dex.
Decreasing the bins size by a factor of a few do not change the results shown here.}.
The shuffling is then implemented in each group separately.
When computing the CF for massive galaxies, we always run 20
different random realizations of shuffling, and plot the average CF
values.

In Fig.~\ref{fig:cf_shuffled_A0} we compare the CFs
of model $A0$ against two basic shuffling tests. When we shuffle between all
galaxies of the same $\mfal,$ the resulting difference in the CF is
very large. It reaches a factor of $\sim4$ in the most massive bin, and a
factor of $\sim2.5$ in the second most massive bin. These large
differences indicate that treating satellite and central galaxies
with the same $\msh$ relation does not agree with SAMs. 
We will show below that the $\msh$ relation for central galaxies is very 
similar to that for satellites in all of the models used here when considering low-mass galaxies.
This is probably the reason why the CFs at the low mass bins do not show a change
after shuffling. In this paper we do not attempt to reproduce the observed
CFs (except for model $A4$ as explained above), but rather compare 
the results of different models. Thus, the
observational data presented in Fig.~\ref{fig:cf_shuffled_A0} are
only meant to provide a basis for comparison between the models. There
is probably no special meaning to the fact that the shuffled samples
better agree with the observed CFs.

Interestingly, the model shows significant change in the CF at the higher
stellar mass bin, even when performing shuffling within $[\mfal$,type$]$. This
difference reaches a factor of $\sim2$ at small scales.  More minor differences
are apparent in the other mass bins, at the level of 15 to 40 per cent. 
These differences indicate that the SAM contains clustering information that depends 
on something other than $\mfal$ and galaxy type.

We have tested the effect of shuffling within $[\mfal$,type$]$ on all the models
listed in Table \ref{tab:models} and found that variations between the different
models are relevant mainly for the most massive stellar mass bin. We therefore
show the CFs for this bin, and for all the models, in
Fig.~\ref{fig:massive_cfs}.  Some models do not show any difference between the
original and the shuffled samples, while other models exhibit substantial
differences. Consequently, model $A0$ is not unique in showing differences in
CFs due to shuffling.  Although model $B0$ does not include any change in the
CFs \citep[in agreement with][]{Wang06}, changing the dynamical friction and
stripping time-scales in model $B2$ introduces a change in the CFs due to
shuffling.

In order to find the reason for the behaviour of the shuffled catalogs we first did the
following tests:
\begin{enumerate}
\item We shuffled only galaxies with one specific type, leaving all other type
  of galaxies unchanged. The results of this test are that only galaxies of type
  1 \& 2 contribute to the shuffling difference. Central (type 0) galaxies do
  not change their clustering properties because of shuffling. We will therefore
  examine below mechanisms which affect the evolution of satellite galaxies.
\item We tested different definitions of the host subhalo mass instead of
  $\mfal$: the maximum mass in the history of the subhalo, and the mass at the
  \emph{first} infall identified by the simulation (our usual definition of
  $\mfal$ is at the \emph{lowest} redshift where the subhalo was identified as
  central, i.e. the \emph{last} infall). All these definitions give rise to the
  same CF sensitivity under shuffling.
\end{enumerate}
In the next section we will discuss the various differences between
SAMs and ABMs in order to better understand the possible reasons for
the effect of shuffling on the CFs.

\section{Differences between ABM \& SAM}
\label{sec:differences}

The results of the shuffling test shown in the previous section
indicate that galaxies produced by ABMs cluster differently than
SAM galaxies. In order to further study this effect we closely follow
each of the assumptions made by ABMs and see if it is fulfilled by our SAMs.
Our motivation is to study second order effects, which are currently not being used by
ABMs, that contribute to the clustering properties of galaxies.

\subsection{Mass function of subhaloes}

\begin{figure}
\centerline{ \hbox{ \epsfig{file=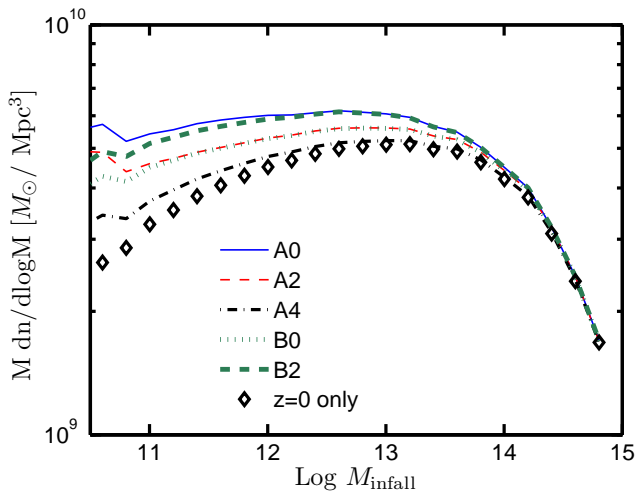,width=9cm} }}
\caption{The abundance of subhaloes with mass $\mfal$, obtained by assigning $\mfal$ for
each galaxy from the SAM catalog at $\z=0$. For type 2 galaxies, we assign $\mfal$
according to the subhalo in which these galaxies were last identified as type 1 or 0.
Different line types correspond to different specific SAMs,
as listed in Table \ref{tab:models}. The diamond symbols show the abundance of
$\mfal$ for subhaloes which are identified at $\z=0$ only.}
  \label{fig:subhalo_mf}
\end{figure}

ABMs often consider subhaloes which are identified at the inspected redshift
only (i.e. at $\z=0$). Although modern high-resolution cosmological simulations
seem to resolve substructure well enough \citep{Conroy06, Moster10, Guo10a},
there is still some uncertainty whether this
resolution is adequate \citep{Guo10}. On the other hand, SAMs follow galaxies even if their
host subhalo has already merged into a bigger one. This is done by allowing
galaxies to survive an additional time inside their 
host halo according to dynamical friction estimates (see
\S\ref{sec:formalism_mergers}).  As a result, the number of galaxies at $\z=0$
in the SAMs might be significantly higher than the number of subhaloes at the
same redshift, where the additional galaxies are all marked as type 2.

In order to quantify this effect we assign $\mfal$ for each galaxy in the SAM
catalog at $\z=0$, as was explained in section \ref{sec:abm_method}. For type 2
galaxies we use $\mfal$ according to the subhalo when the galaxy was last identified as type
0 or 1. The resulting mass functions for all the models are plotted in
Fig.~\ref{fig:subhalo_mf}.
It is evident that all of the SAMs contain a non-negligible number of type 2
galaxies at $\z=0$.
In models that use a long dynamical friction time scale, the
difference is bigger, and can be significant even at masses that are
two orders of magnitude above the minimum mass of the
simulation\footnote{Various ABMs take this effect into consideration
and add subhaloes from high redshift according to dynamical friction
estimates, see e.g. \citet{Moster10}.}.

Although the difference in the number of galaxies is sometimes small in
comparison to the total population, these type 2 galaxies might still affect the
CF.  In particular, such galaxies are usually located close to the central galaxies of their
\fof group, and thus contribute to the CF at small scales, where a small number of
galaxies can make a significant difference. The effect can be seen by comparing
the CF of model $A0$ against model $A4$ (Fig.~\ref{fig:massive_cfs}). These
models mainly differ in the way dynamical friction is modeled, resulting in a
significant change in the CF at small scales.

The shuffling test done here does not change the number of type 2
galaxies so different
mass functions of $\mfal$ do not affect this test directly. This can be seen when
comparing Fig.~\ref{fig:massive_cfs} and \ref{fig:subhalo_mf}, there is no
obvious correlation between models that are sensitive to shuffling and models
with high abundance of $\mfal$.

\subsection{Redshift dependence}

\begin{figure*}
\hbox{ \epsfig{file=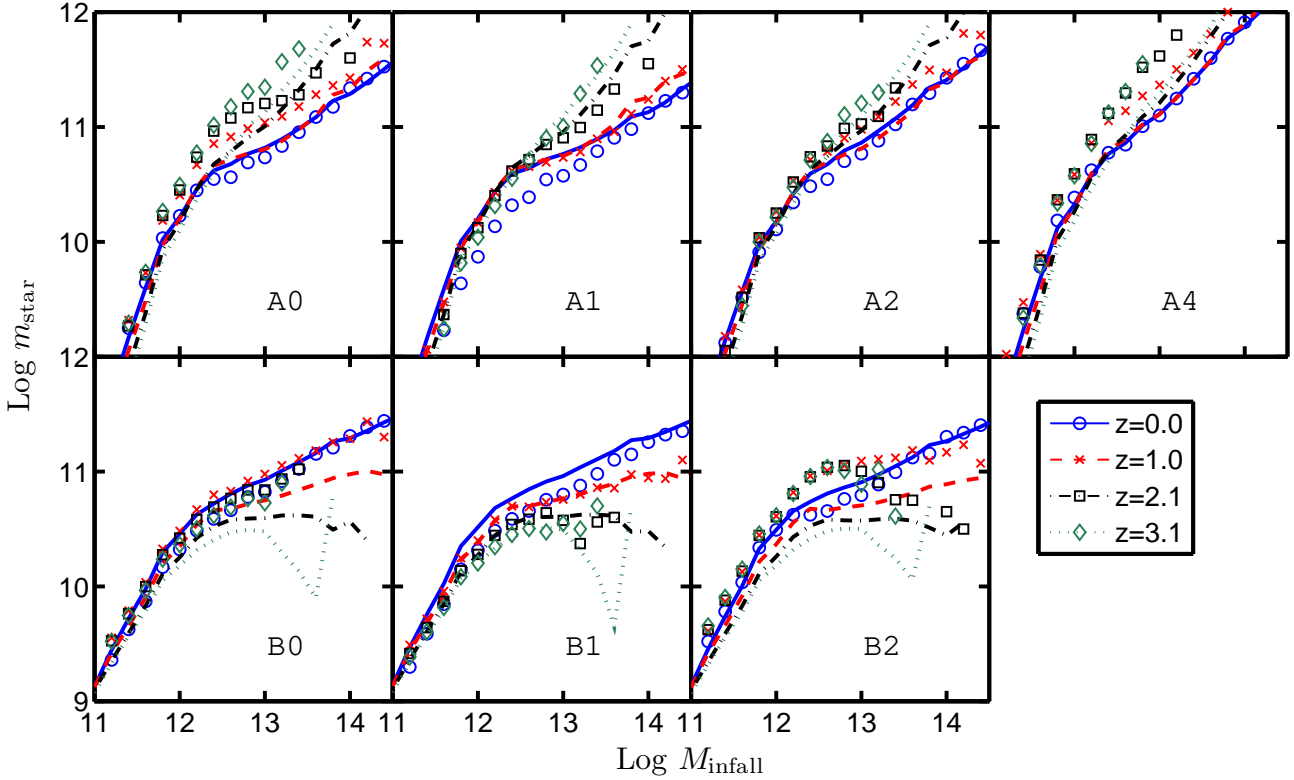,width=18cm} }
\caption{The $\msh$ relation for the various models
used in this work and for different redshifts. Each panel shows results
from a specific model as listed in Table \ref{tab:models}.
All the values of $\ms$ were derived by computing the average per a given $\mfal$ mass bin.
\emph{Lines} show $\ms$ for type 0 galaxies which are selected at the indicated redshift.
\emph{Symbols} show $\ms$ for type 1 \& 2 galaxies identified at $\z=0$, split according to their infall
redshift $\zfal$. For type 1 \& 2 galaxies, we allow $\zfal$ to deviate in $\Delta\z=\pm0.2$
from the indicated $\z$ label ($\Delta\z=\pm0.5$ for $\z=3$).}
\label{fig:ms_mh_all1}
\end{figure*}

In Fig.~\ref{fig:ms_mh_all1} we show the $\msh$ relation for the
SAMs used here for various redshifts.
Clearly, the relation between stellar mass and subhalo mass depends
on redshift for central galaxies, as was found by previous ABM studies
\citep[e.g.,][]{Conroy09}. This dependence is mainly related to the
evolution of the stellar mass functions with redshift.
At the massive end, the average $\msh$ relation is also affected by
the scatter in $\ms$ for a given value of $\mfal$. For low
mass galaxies, all the models show little or no evolution with
redshift, in contrast to the relation derived by \citet{Conroy09}.
This might be a consequence of the different stellar mass functions
used by these authors.
A different important point to note is that the $A$ models show very different
behaviour than the $B$ models.  This difference is an outcome of the stellar
mass function and its evolution with redshift. The $B$ models produce too many
small mass and high mass galaxies, so their $\msh$ relations are probably incorrect.

The dependence on redshift, which is rather obvious for central galaxies (smooth
lines in Fig.~\ref{fig:ms_mh_all1}), is also important for \emph{satellite}
galaxies identified at $\z=0$. In Fig.~\ref{fig:ms_mh_all1} we plot the average
$\ms$ for satellite galaxies with different $\zfal$ as symbols. In the $A$ models,
galaxies with higher infall redshift have higher current stellar mass, while
the $B$ models show the opposite trend. This effect is not included in current ABMs
when matching the $\msh$ relation at redshift zero. Shuffling between satellite
galaxies of the same $\mfal$ will mix galaxies of different
$\zfal$, an effect that might change the CF and contribute to the differences
shown in Fig.~\ref{fig:massive_cfs}.

We note that the average $\ms$ per a given $\zfal$ seen in
Fig.~\ref{fig:ms_mh_all1} does not provide enough information
in order to quantify the effect of
shuffling, as it does not include information on the \emph{number}
of galaxies per each $\zfal$. The population of satellite galaxies
is dominated by low values of $\zfal$, so only values of $\zfal\lesssim1$
are relevant. This issue will be further examined in section \ref{sec:add_param}.

\subsection{Evolution at early times}
\label{sec:early_diff}

In the previous section we discussed the correlation between $\zfal$ and $\ms$
for satellite galaxies, and for a given $\mfal$. However, the average $\ms$ for
satellite galaxies at $\z=0$ with a specific $\zfal=\z_0$ is different from the $\ms$ of
central galaxies at the same $\z_0$. This is seen by comparing lines and symbols in
Fig.~\ref{fig:ms_mh_all1}. As a result, one cannot use the $\msh$ relation
that was derived for central galaxies at high-redshift \citep[e.g.,][]{Conroy09},
in order to model the $\msh$ for satellites at $\z=0$ with various $\zfal$.
Here we will explore one reason for this effect, namely the difference
in $\ms$ at the time of infall. In the next section we will study the growth
in $\ms$ after $\zfal$.

For a distinct \fof halo, it is well established that the
large-scale over-density of its environment is correlated with the halo formation
history. According to this `environmental effect', haloes of a given mass living in denser
environments are formed earlier \citep[this is also termed the `assembly bias':][]{Gao05,Harker06,Wechsler06}.
The effect introduces a weak correlation between the stellar mass of galaxies in the SAMs
and the environment of their host \fof halo \citep{Croton07}.
As galaxies identified just before $\zfal$ live in denser
environments than the average central galaxy,
this effect might generate a difference in $\ms$ between all central galaxies
and future satellite galaxies. Moreover, for satellite galaxies that fall into
a larger halo at high-$\z$, the environment is correlated to the mass of the
target \fof group at $\z=0$. Thus, we expect that $\ms$ will be correlated with
the group mass $\mf$ and not only with $\mfal$.

\begin{figure*}
\centerline{ \hbox{ \epsfig{file=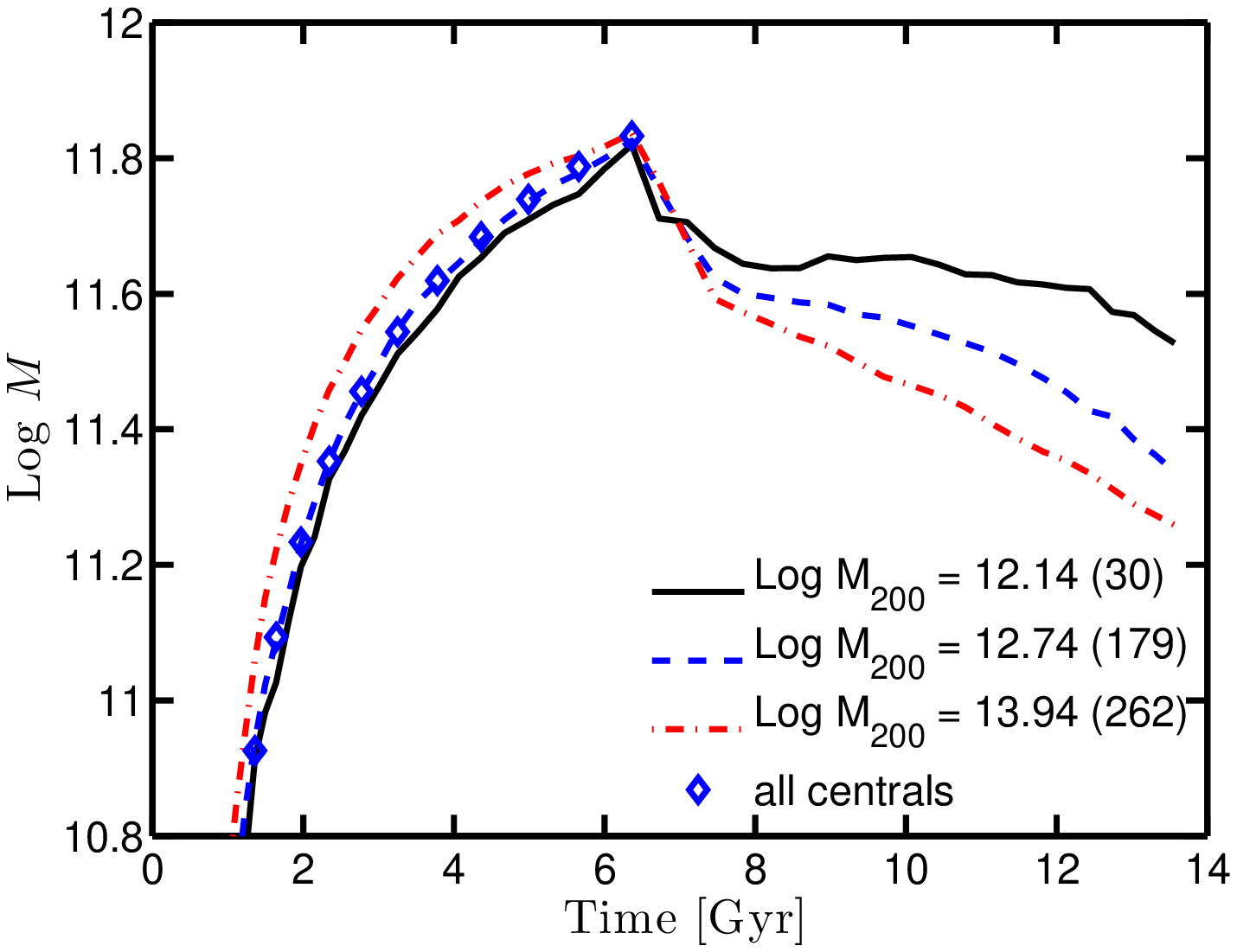,width=9cm}
\epsfig{file=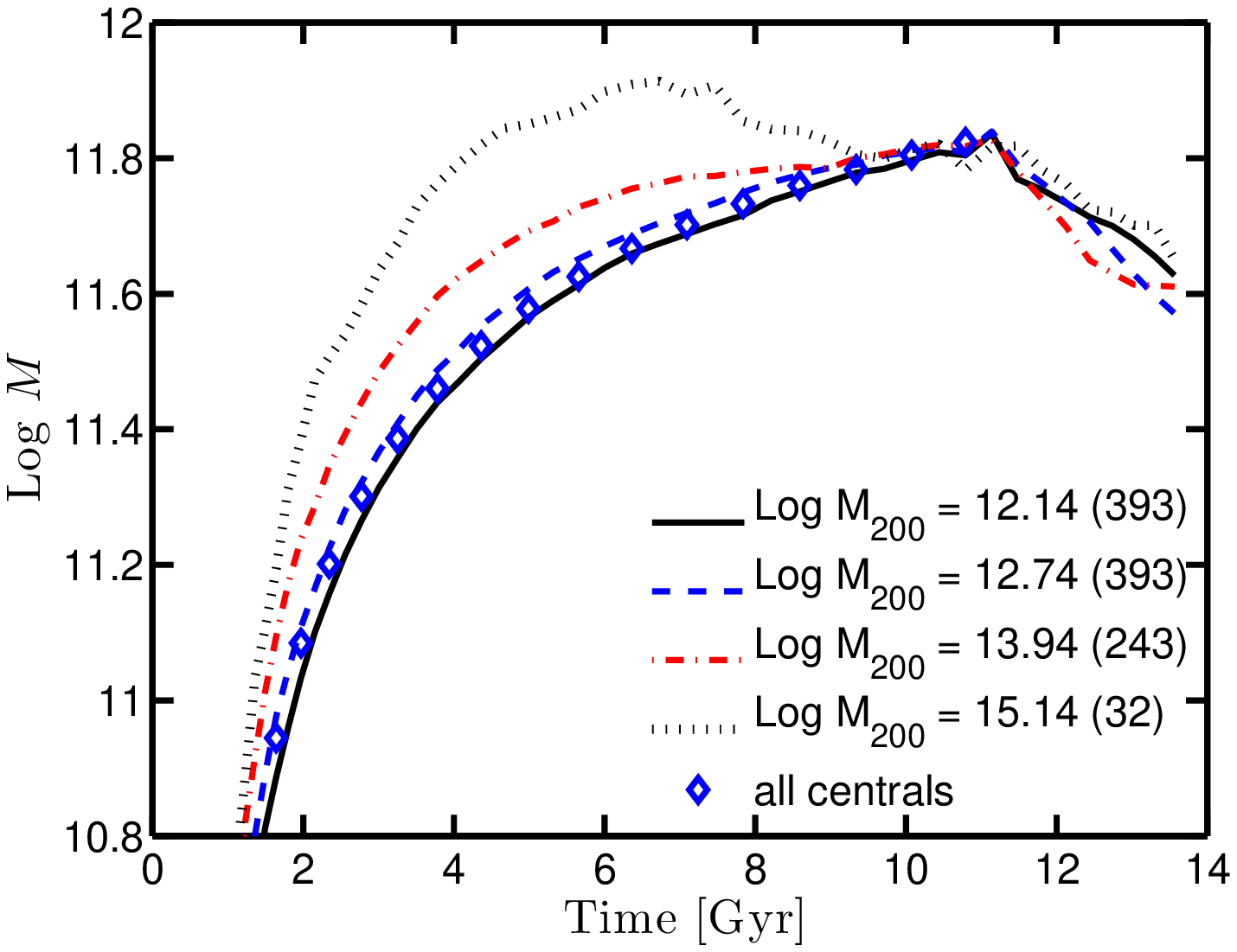,width=9cm} }}
\caption{Main-progenitor histories of subhaloes with the same infall
mass and redshift, split by their host mass at $\z=0$, $\mf$.
Each panel includes histories for a given values of $\zfal$ and $\mfal$.
We track the main-progenitor histories starting $\z=0$, so only
subhaloes which survive until $\z=0$ are included. Diamond symbols show
the averaged main-progenitor histories for all central subhaloes
which have the same mass at the infall redshift. Each $\mf$ bin
includes values within 0.2 dex in Log$\msun$, where the number of main-progenitor
histories used in each bin is given in parentheses. All values are taken
from the full Millennium simulation. Time is in Gyr since the big-bang.}
  \label{fig:subhalo_mp}
\end{figure*}

We first test this effect for the underlying dark-matter evolution, using
\emph{subhaloes} from the Millennium simulation. In Fig.~\ref{fig:subhalo_mp} we
show the main-progenitor
histories of subhaloes with the same $\mfal$ and $\zfal$. We split the average
histories into different subsets of equal $\mf$ (the host of these subhaloes at
$\z=0$).  Also plotted in the same figure are the average histories of central
subhaloes identified at the same $\zfal$. Indeed, merger-histories at early
times are correlated to the $\mf$ mass at $\z=0$, especially if the infall
redshift is low.  This effect is very similar to the environmental effect
mentioned above, where subhaloes that fall into more massive $\mf$ form earlier
(i.e., the redshift which corresponds to half the current mass is higher).

In order to test the effect of $\mf$ on early formation histories of
\emph{galaxies}, we stop all modes of SF and merging for satellite galaxies in
models $A1$ and $B1$. In all other aspects these models are the same as models
$A0$ and $B0$. Consequently, in models $A1$ and $B1$, the stellar mass of
satellite galaxies at $\z=0$ is identical to their mass at $\zfal$.  The reader
is referred again to Fig.~\ref{fig:ms_mh_all1}, where the $\msh$ relation is
plotted for models $A1$ and $B1$. Not surprisingly, the stellar mass of
satellite galaxies is already different at $\zfal$ from the stellar mass of
central galaxies. This difference can reach 70 per cent in model $A1$ for the
average value of $\ms$, at $\mfal\sim3\times10^{12}\msun$.

The effect of $\mf$ on $\ms$ at early epochs can modify the CFs, and might
contribute to the effect of shuffling discussed in section \ref{sec:shuffle}.
The reason that there is no change in the CF due to shuffling in models $A1$ and 
$B1$ might be that the effect of environment is maximal in these models, which seems 
to cancel the contribution of other effects going in the opposite direction.
We will discuss this issue further in section \ref{sec:add_param}.

\subsection{Stellar mass growth after infall}

The amount of stellar mass gained by galaxies of type 1 \& 2 after $\zfal$ can be
seen in Fig.~\ref{fig:ms_mh_all1}, when comparing the various models to models $A1$,$B1$.
Models with slow gas stripping within satellite galaxies ($A0$, $A4$ \& $B2$) show
a significant
increase of mass after infall, while models with fast stripping show
only a minor change ($B0$ \& $A2$).  In models with slow stripping mechanism, more hot gas is available
for cooling, and consequently more cold gas can reach the disk and form stars.
Both observational and theoretical studies indicate that slow stripping
is the preferred scenario for modeling the environmental evolution of galaxies
within their host halo \citep{McCarthy08,Font08,Khochfar08,Weinmann10}.

In model $A4$ we use stripping of hot gas which is proportional to the dark-matter
stripping. This scenario was found by \citet{Weinmann10} to reproduce the fraction 
of passive satellite galaxies as a function of group mass and distance
from the group center. In this case, the stripping rate will be correlated to the
dark-matter evolution. From Fig.~\ref{fig:subhalo_mp} it seems that the dark-matter
evolution of subhaloes after $\zfal$ is affected by $\mf$, the group mass at $\z=0$.
This is probably a natural consequence of the tidal interaction within the group.
As a result of the above, the amount of $\ms$ gained after infall should correlate
with $\mf$ in this model. On the other hand, it may be expected that galaxies with
higher $\zfal$ will have more time to increase their $\ms$, so correlation with $\zfal$
is also expected.

\begin{figure}
\centerline{ \hbox{ \epsfig{file=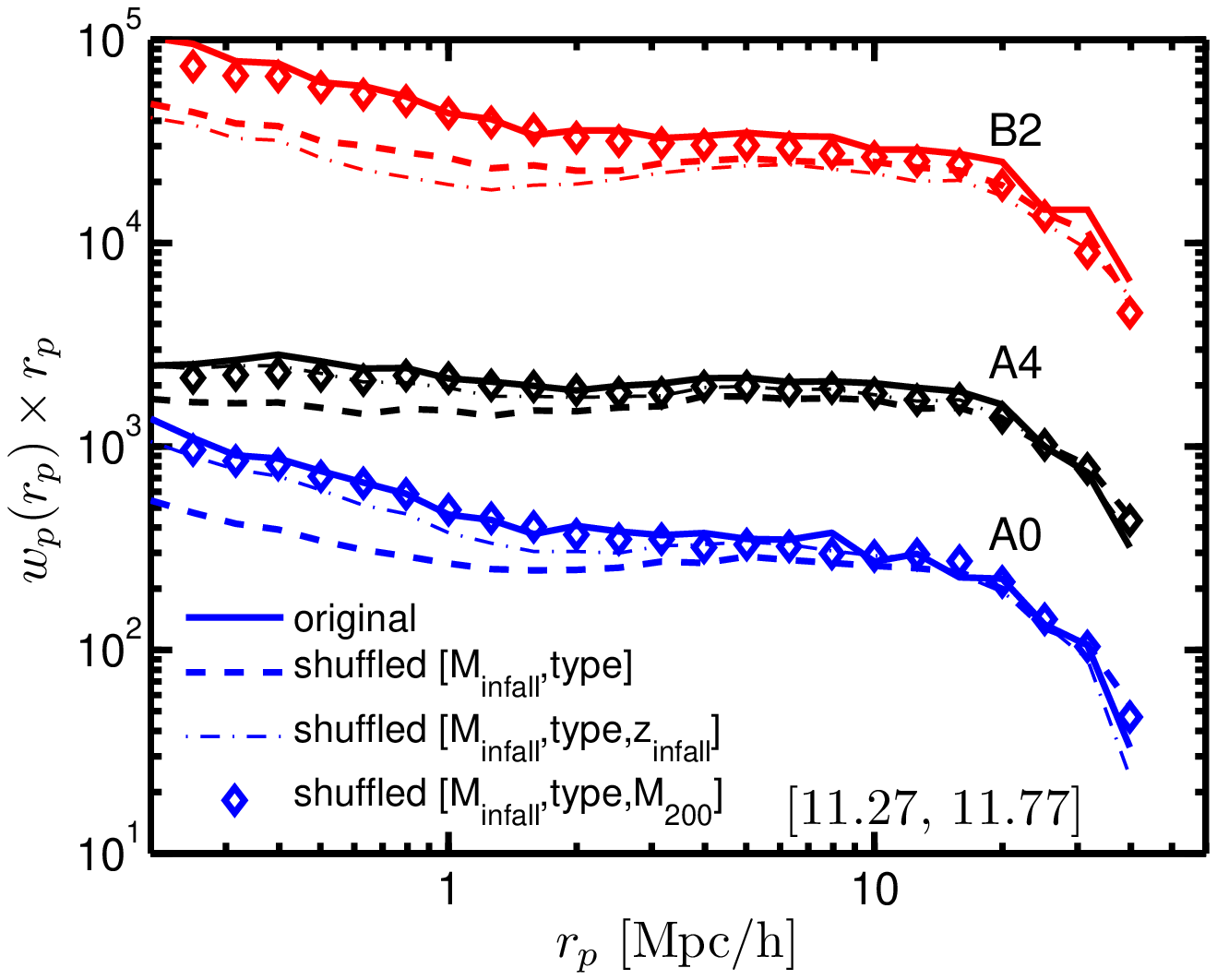,width=9cm} }}
\caption{The CFs for various models using a range in $\ms$ as indicated in units of Log$\msun$.
For each model we run a few different shuffling tests, where the groups of shuffled
galaxies are constrained to have the same set of properties:
$\mfal$ and galaxy type (dashed lines);
$\mfal$, galaxy type, and $\zfal$ (dashed-dotted lines);
$\mfal$, galaxy type, and $\mf$ (diamonds).
An ABM model using two parameters for fixing the stellar mass of galaxies, 
namely $\mfal$ and $\mf$, is shown to reproduce the CFs of the SAM galaxies.}
 \label{fig:shuff1}
\end{figure}

\section{ABM with an additional parameter}
\label{sec:add_param}

\begin{figure*}
\centerline{ \hbox{ \epsfig{file=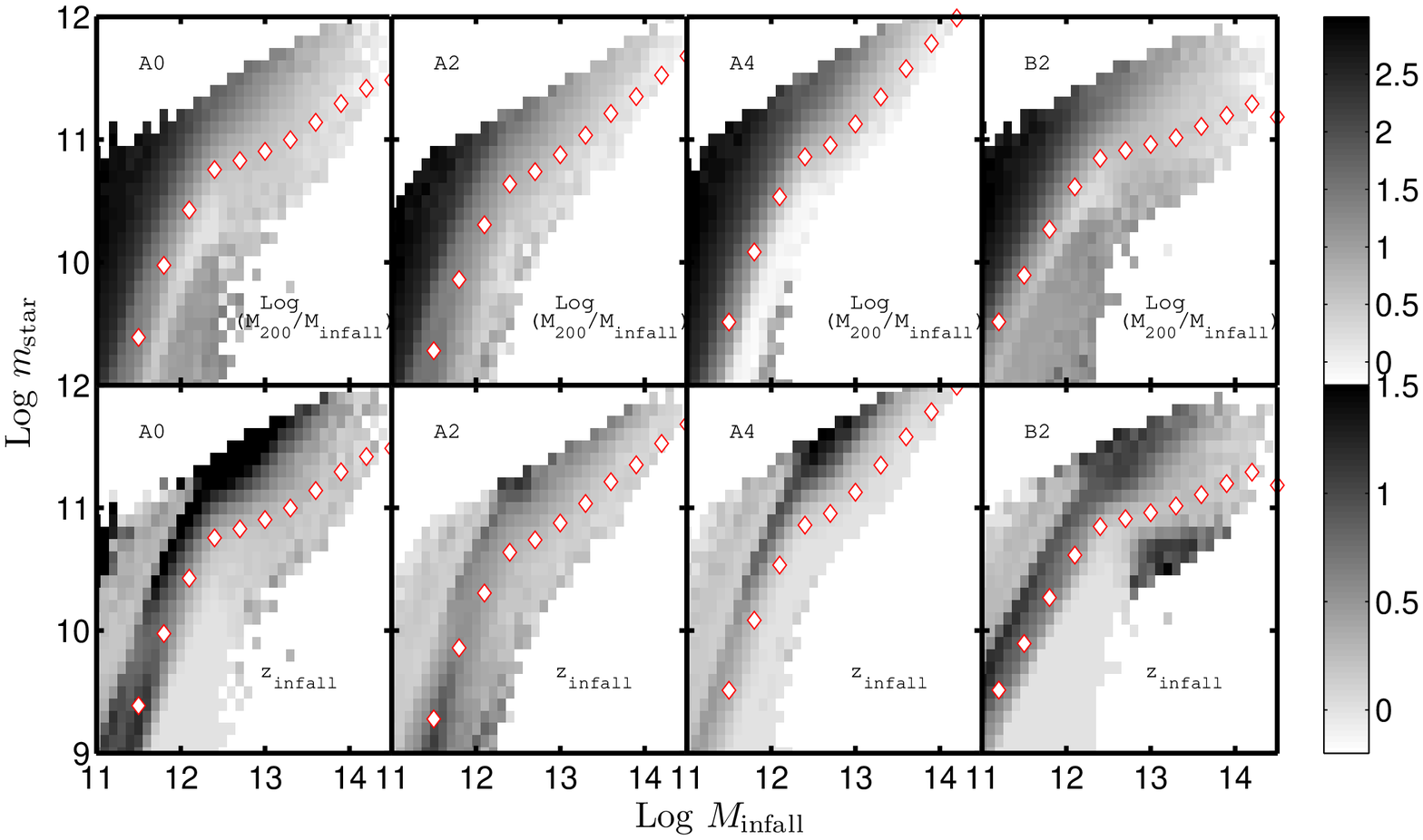,width=18cm} }}
\caption{The behaviour of $\mf$ and $\zfal$ as a function of $\mfal$
and $\ms$ for satellite galaxies (type 1 \& 2).
\emph{Upper panels:} median values of Log $\mf/\mfal$;
\emph{Lower panels:} median values $\zfal$.
Color coding is set to span 95 per cent of the population of $\zfal$ and $\mf/\mfal$.
Each panel shows results for one model as indicated. For the list of all models
please see Table \ref{tab:models}.
Median values are computed for bins with more than 10 galaxies. Diamonds show the
median $\ms$ for all galaxies in the same $\mfal$ bin.}
\label{fig:3d_dep}
\end{figure*}

In section \ref{sec:shuffle} we saw that SAMs include additional clustering
information which is not reproduced by a dependence on $\mfal$ only. This was
demonstrated by the different CF results when performing shuffling. We then learned
that a single $\msh$ relation cannot reproduce accurately the
behaviour of satellite galaxies. It is not easy to disentangle the different
reasons for the deviation in the $\msh$ relation, and their affect on the CFs. This will
require studying SAMs which include only one deviation per a model. For example, we would need
to develop a SAM in which there is no redshift evolution in the mass relation, but satellite
galaxies do increase their stellar mass function after infall.
Here we choose to present a more modest test. Instead of modifying the SAMs, we will present
a modification of ABMs which is able mimic the behaviour of SAMs better.

The target of this section is to present a more detailed version of ABMs which is
relatively simple to use, and is able to fit the CF of SAMs.
This is achieved by adding an additional parameter to ABMs, so the stellar mass of a galaxy
is fixed not only by $\mfal$ and the galaxy type, but also according to an additional
ingredient. According to the previous section, there are two parameters which naturally
affect the stellar mass of satellite galaxies: the infall redshift ($\zfal$), and the
group mass at redshift zero ($\mf$). These parameters are correlated
because of the hierarchical build-up of haloes (galaxies inside more massive
groups tend to have a higher $\zfal$). Consequently, it is hard to separate the
effect of each parameter on the models.

An important criterion for choosing the second parameter is the behaviour of the shuffled CFs.
In order to check this issue, we present two additional shuffling tests. For each of the parameters
tested ($p=\zfal$,$\mf$) we divide the catalog of galaxies into groups of the same $\mfal$, type, and
$p$ values as discussed in section \ref{sec:shuffle}. The location of galaxies are shuffled within
these new groups. The resulting CFs are shown in Fig.~\ref{fig:shuff1}. For all the models tested here,
using the group mass, $\mf$, results in no change in the CF after shuffling. This means that SAMs have no additional clustering
information except the dependence on $\mfal$, $\mf$, and galaxy type\footnote{According to \citet{Croton07}
there are deviations on the order of 10 per cent because of environmental effects on scales larger than the halo.
These variations are too small to be seen in Fig.~\ref{fig:shuff1}.}. Consequently, ABMs that would use
$\mf$ to fix the stellar mass in addition to $\mfal$ would be able to reproduce the results
of SAMs. Such models might have more freedom and could span more possible $\msh$ relations.

Naively, one may expect that the main variation in the $\msh$ relation should
be due to the the additional dependence of satellite mass on $\zfal$.
However, this parameter does not allow ABMs to better
follow the CF of SAMs. In Fig.~\ref{fig:shuff1}, adding $\zfal$ as a second parameter makes
the CF of model $B2$ to deviate even more from the original CF. Adding $\zfal$ to model $A0$
improves the CF, but it does not reach the accuracy achieved by using $\mf$.
The failure of $\zfal$ to improve ABMs is surprising, as it naturally affects the stellar
mass of satellite galaxies. However, it seems that its correlation with stellar mass is weaker
than the correlation between $\mf$ and stellar mass.

We have examined various statistical aspects of the effect of $\mf$
and $\zfal$ on $\ms$ (per a given $\mfal$). The main obstacle in
presenting a clear-cut evidence for why $\mf$ works better than $\zfal$ is the
different abundance of satellite galaxies as a function of $\zfal$
and $\mf$. For example, most satellite galaxies have $\zfal<1$ \citep[see e.g.][]{Gao04},
but have a different unique range in $\mf/\mfal$. A clear
comparison should take this number distribution into account.
We therefore plot in Fig.~\ref{fig:3d_dep} the median values of $\mf$ and $\zfal$
per each bin in $\ms$ and $\mfal$. When comparing $\mf$ against
$\zfal$, the number of galaxies per each bin remains the same. However,
we note that most of the galaxies are concentrated around the median
$\ms$ values (diamond symbols).

Fig.~\ref{fig:3d_dep} shows the reason for the success of $\mf$ as a second parameter. The dependence of $\ms$
on $\mf$ (for a given $\mfal$) is roughly monotonic and smooth. For a given $\mfal$, the values
of $\mf$ span nicely the scatter in $\ms$. A different behaviour is seen for $\zfal$.
The relation between $\ms$ and $\zfal$ for a given $\mfal$ is not monotonic. This means that
the dependence of $\ms$ on $\zfal$ is in general not unique, so very different $\ms$ correspond
to the same $\zfal$. 
For example, consider model $B2$ (lower-right panel): shuffling galaxies with the same $\mfal\sim10^{13}$
and $\zfal\sim1$ will result in swapping galaxies between very different stellar masses (the dark regions). 
As can be seen from the upper panel, the swapped galaxies live inside 
different \fof groups, so this shuffling will affect the CF significantly.
This is probably the reason why $\zfal$ does not improve the
shuffled CF in model $B2$. We note that the range in $\ms$ used for
testing the shuffled CFs in Fig.~\ref{fig:shuff1} corresponds to
$\mfal\gtrsim10^{12}\msun$. As the number of subhaloes of a given $\mfal$ decreases quickly
with increasing $\mfal$; the CF in this mass range is mostly affected by $\mfal\sim10^{12}\msun$.

\section{Summary and discussion}
\label{sec:discuss}

In this work we have studied the relation between the mass of subhaloes
($\mfal$) and the stellar mass of galaxies. A simple model for
this relation is crucial for summarizing the basic concepts of
galaxy formation physics. It can then be used to model the correlation
function of galaxies, their stellar mass function, star-formation rates,
and merging processes. The abundance matching approach (ABM) is aiming at
constraining this mass relation directly, without the need to rely on
the assumptions of a specific model. This constraint is important
because other theoretical methodologies depend strongly on the detailed physics
assumed, and are usually not able to match the observational constraints
accurately. It is therefore necessary to study the assumptions made by the 
ABM approach, and their effect on the mass relation above.

In this paper we have tested the basic assumptions of ABM models using a
set of Semi-Analytical Models (SAMs) developed by \citet{Neistein10}.
We examined the relation between stellar mass and $\mfal$ within the
SAMs at different redshifts, and for satellite versus central galaxies.
It was shown that a single $\msh$ relation is not able to capture
accurately the full complexity of SAMs, due to
complex behaviour of satellite galaxies. First, the stellar mass
of these galaxies behaves differently at early times due to {an
environmental effect acting on their host haloes}
\citep[the `assembly bias';][]{Gao05,Harker06}.
Satellite galaxies falling into more massive groups
have different dark-matter merger-histories, and are more massive even
before falling into their group. Second, the mass of satellite galaxies
depends on the infall redshift ($\zfal$, defined as the time they became satellites).
This is because they roughly follow the global behaviour of central galaxies.
Third, satellite galaxies acquire a significant amount of stellar mass after
$\zfal$, especially if the model assumes slow gas stripping mechanisms.
This growth can reach a factor of $\sim2$ in stellar mass, for
galaxies with $\zfal=1$.

We have used the shuffling technique, where the location of all
model galaxies that have the same $\mfal$ and galaxy type (central/satellite)
are swapped randomly. Assuming that the stellar mass of galaxies depends only on the
host subhalo mass ($\mfal$), is equivalent to using a shuffled sample of galaxies.
In order to check the effect of this assumption on the clustering properties
of galaxies, we compared the auto correlation
functions (CFs) of the original SAMs against the shuffled catalogs.
We found that shuffled catalogs of galaxies can have different
CFs, depending on the specific model being tested. The shuffled CFs
are very different if we assume the same $\msh$ relation for
satellite and central galaxies, reaching a factor of $\sim4$ at
scales below 1 Mpc. However, even when using a different $\msh$ relation
for central and satellite galaxies, this difference can reach a
factor of 2 in the CF of the most massive galaxies.

The results shown here are based on a few specific SAMs and might
not reflect the `true' physical Universe. It might be that the
assumptions adopted by ABM models are correct, and SAMs introduce more
complexity than what is needed. However, it might be that the `true'
model includes a different set of assumptions, or is more extreme in
violating the assumptions tested here. A different
limitation of this work is that
the results of the auto-correlation functions
obtained here are based on the Millennium simulation. This
simulation assumes $\sigma_8=0.9$, which is higher than the latest
favorable estimates. As a result, the relation
between the model CFs and the observed data is of less interest to
this study. We did not aim at providing a model which reproduce the observed
values of both the stellar mass function and the CFs.

We have checked two additional parameters which could be adopted by
future ABM models in order to better reproduce the results of SAMs. For the
SAMs used here, assuming that the stellar mass depends on $\mfal,\,\mf$
and galaxy type is enough to reproduce the CF of galaxies ($\mf$ is
the \fof group mass at redshift zero). The other natural parameter,
$\zfal$ does not provide a good match to the clustering of SAMs
probably because it is less correlated with the stellar
mass. Using $\mf$ as a second parameter makes ABM models more similar
to HODs, in which all the properties of galaxies are fixed by the group mass.

Our results may be relevant for interpreting the 
observed environmental dependencies of galaxies.
While various results
stress the importance of late evolution on the properties
of satellite galaxies \citep[e.g.][]{Kauffmann04},
we find evidence for the influence of early evolution. At a fixed $\mfal$,
galaxies ending up in massive clusters undergo a more rapid growth of their dark
matter halo
at early times, which may influence their observed properties at late times, like
their morphology or star formation rate.

Studies like \citet{vdBosch08} and \citet{Weinmann09} investigate the impact of environment by comparing satellite and central
galaxies at fixed stellar mass today. \citet{vdBosch08} argue
that this is reasonable, as most satellites fall in late, and thus no large difference in the evolution
of stellar mass between central and satellite galaxies since this time
is expected. However,
as we have shown here, the progenitors of today's satellites and the progenitors of today's centrals may be different
already at redshifts before the time of infall. Depending on the exact nature
and magnitude of this effect, it may complicate the
direct comparison between central and satellite galaxies.
In a future work we intend to study the differences in stellar mass between
satellite and central galaxies in more detail, making use of the
two parameter ABM approach suggested in this work.


\section*{Acknowledgments}

We thank Simon White and Sadegh Khochfar for useful discussions, and
for a careful reading of an earlier version of this manuscript.
EN thanks the department of astrophysics in Tel-Aviv University,
for kindly hosting him while working on this project.
EN is supported by the Minerva fellowship.

\bibliographystyle{mn2e}
\bibliography{ref_list}

\begin{thebibliography}{}

\bibitem[\protect\citeauthoryear{{Baldry}, {Glazebrook} \& {Driver}}{{Baldry}
  et~al.}{2008}]{Baldry08}
{Baldry} I.~K.,  {Glazebrook} K.,    {Driver} S.~P.,  2008, \mnras, 388, 945

\bibitem[\protect\citeauthoryear{{Behroozi}, {Conroy} \& {Wechsler}}{{Behroozi}
  et~al.}{2010}]{Behroozi10}
{Behroozi} P.~S.,  {Conroy} C.,    {Wechsler} R.~H.,  2010, \apj, 717, 379

\bibitem[\protect\citeauthoryear{{Berlind} \& {Weinberg}}{{Berlind} \&
  {Weinberg}}{2002}]{Berlind02}
{Berlind} A.~A.,  {Weinberg} D.~H.,  2002, \apj, 575, 587

\bibitem[\protect\citeauthoryear{{Bernardi}, {Shankar}, {Hyde}, {Mei},
  {Marulli} \& {Sheth}}{{Bernardi} et~al.}{2010}]{Bernardi10}
{Bernardi} M.,  {Shankar} F.,  {Hyde} J.~B.,  {Mei} S.,  {Marulli} F.,
  {Sheth} R.~K.,  2010, \mnras, 404, 2087

\bibitem[\protect\citeauthoryear{{Binney} \& {Tremaine}}{{Binney} \&
  {Tremaine}}{1987}]{Binney87}
{Binney} J.,  {Tremaine} S.,  1987, {Galactic dynamics}

\bibitem[\protect\citeauthoryear{{Bode}, {Ostriker} \& {Turok}}{{Bode}
  et~al.}{2001}]{Bode01}
{Bode} P.,  {Ostriker} J.~P.,    {Turok} N.,  2001, \apj, 556, 93

\bibitem[\protect\citeauthoryear{Borch et~al.,}{Borch  et~al.}{2006}]{Borch06}
Borch A.,  et~al., 2006, \aap, 453, 869

\bibitem[\protect\citeauthoryear{Bundy et~al.,}{Bundy  et~al.}{2006}]{Bundy06}
Bundy K.,  et~al., 2006, \apj, 651, 120

\bibitem[\protect\citeauthoryear{{Conroy}, {Prada}, {Newman}, {Croton}, {Coil},
  {Conselice}, {Cooper}, {Davis}, {Faber}, {Gerke}, {Guhathakurta}, {Klypin},
  {Koo} \& {Yan}}{{Conroy} et~al.}{2007}]{Conroy07}
{Conroy} C.,  {Prada} F.,  {Newman} J.~A.,  {Croton} D.,  {Coil} A.~L.,
  {Conselice} C.~J.,  {Cooper} M.~C.,  {Davis} M.,  {Faber} S.~M.,  {Gerke}
  B.~F.,  {Guhathakurta} P.,  {Klypin} A.,  {Koo} D.~C.,    {Yan} R.,  2007,
  \apj, 654, 153

\bibitem[\protect\citeauthoryear{{Conroy} \& {Wechsler}}{{Conroy} \&
  {Wechsler}}{2009}]{Conroy09}
{Conroy} C.,  {Wechsler} R.~H.,  2009, \apj, 696, 620

\bibitem[\protect\citeauthoryear{{Conroy}, {Wechsler} \& {Kravtsov}}{{Conroy}
  et~al.}{2006}]{Conroy06}
{Conroy} C.,  {Wechsler} R.~H.,    {Kravtsov} A.~V.,  2006, \apj, 647, 201

\bibitem[\protect\citeauthoryear{{Cox}, {Jonsson}, {Somerville}, {Primack} \&
  {Dekel}}{{Cox} et~al.}{2008}]{Cox08}
{Cox} T.~J.,  {Jonsson} P.,  {Somerville} R.~S.,  {Primack} J.~R.,    {Dekel}
  A.,  2008, \mnras, 384, 386

\bibitem[\protect\citeauthoryear{Croton et~al.,}{Croton
  et~al.}{2006}]{Croton06}
Croton D.~J.,  et~al., 2006, \mnras, 365, 11

\bibitem[\protect\citeauthoryear{{Croton}, {Gao} \& {White}}{{Croton}
  et~al.}{2007}]{Croton07}
{Croton} D.~J.,  {Gao} L.,    {White} S.~D.~M.,  2007, \mnras, 374, 1303

\bibitem[\protect\citeauthoryear{{Davis}, {Efstathiou}, {Frenk} \&
  {White}}{{Davis} et~al.}{1985}]{Davis85}
{Davis} M.,  {Efstathiou} G.,  {Frenk} C.~S.,    {White} S.~D.~M.,  1985, \apj,
  292, 371

\bibitem[\protect\citeauthoryear{{De Lucia} \& {Blaizot}}{{De Lucia} \&
  {Blaizot}}{2007}]{DeLucia07}
{De Lucia} G.,  {Blaizot} J.,  2007, \mnras, 375, 2

\bibitem[\protect\citeauthoryear{{Drory}, {Bender}, {Feulner}, {Hopp},
  {Maraston}, {Snigula} \& {Hill}}{{Drory} et~al.}{2004}]{Drory04}
{Drory} N.,  {Bender} R.,  {Feulner} G.,  {Hopp} U.,  {Maraston} C.,  {Snigula}
  J.,    {Hill} G.~J.,  2004, \apj, 608, 742

\bibitem[\protect\citeauthoryear{{Drory}, {Salvato}, {Gabasch}, {Bender},
  {Hopp}, {Feulner} \& {Pannella}}{{Drory} et~al.}{2005}]{Drory05}
{Drory} N.,  {Salvato} M.,  {Gabasch} A.,  {Bender} R.,  {Hopp} U.,  {Feulner}
  G.,    {Pannella} M.,  2005, \apjl, 619, L131

\bibitem[\protect\citeauthoryear{{Font}, {Bower}, {McCarthy}, {Benson},
  {Frenk}, {Helly}, {Lacey}, {Baugh} \& {Cole}}{{Font} et~al.}{2008}]{Font08}
{Font} A.~S.,  {Bower} R.~G.,  {McCarthy} I.~G.,  {Benson} A.~J.,  {Frenk}
  C.~S.,  {Helly} J.~C.,  {Lacey} C.~G.,  {Baugh} C.~M.,    {Cole} S.,  2008,
  \mnras, 389, 1619

\bibitem[\protect\citeauthoryear{{Fontana}, {Salimbeni}, {Grazian},
  {Giallongo}, {Pentericci}, {Nonino}, {Fontanot}, {Menci}, {Monaco},
  {Cristiani}, {Vanzella}, {de Santis} \& {Gallozzi}}{{Fontana}
  et~al.}{2006}]{Fontana06}
{Fontana} A.,  {Salimbeni} S.,  {Grazian} A.,  {Giallongo} E.,  {Pentericci}
  L.,  {Nonino} M.,  {Fontanot} F.,  {Menci} N.,  {Monaco} P.,  {Cristiani} S.,
   {Vanzella} E.,  {de Santis} C.,    {Gallozzi} S.,  2006, \aap, 459, 745

\bibitem[\protect\citeauthoryear{{Gao}, {Springel} \& {White}}{{Gao}
  et~al.}{2005}]{Gao05}
{Gao} L.,  {Springel} V.,    {White} S.~D.~M.,  2005, \mnras, 363, L66

\bibitem[\protect\citeauthoryear{{Gao}, {White}, {Jenkins}, {Stoehr} \&
  {Springel}}{{Gao} et~al.}{2004}]{Gao04}
{Gao} L.,  {White} S.~D.~M.,  {Jenkins} A.,  {Stoehr} F.,    {Springel} V.,
  2004, \mnras, 355, 819

\bibitem[\protect\citeauthoryear{{Guo}, {White}, {Boylan-Kolchin}, {De Lucia},
  {Kauffmann}, {Lemson}, {Li}, {Springel} \& {Weinmann}}{{Guo}
  et~al.}{2010}]{Guo10}
{Guo} Q.,  {White} S.,  {Boylan-Kolchin} M.,  {De Lucia} G.,  {Kauffmann} G.,
  {Lemson} G.,  {Li} C.,  {Springel} V.,    {Weinmann} S.,  2010, ArXiv
  e-prints: 1006.0106

\bibitem[\protect\citeauthoryear{{Guo}, {White}, {Li} \&
  {Boylan-Kolchin}}{{Guo} et~al.}{2010}]{Guo10a}
{Guo} Q.,  {White} S.,  {Li} C.,    {Boylan-Kolchin} M.,  2010, \mnras, 404,
  1111

\bibitem[\protect\citeauthoryear{{Harker}, {Cole}, {Helly}, {Frenk} \&
  {Jenkins}}{{Harker} et~al.}{2006}]{Harker06}
{Harker} G.,  {Cole} S.,  {Helly} J.,  {Frenk} C.,    {Jenkins} A.,  2006,
  \mnras, 367, 1039

\bibitem[\protect\citeauthoryear{{Kauffmann}, {White}, {Heckman}, {M{\'e}nard},
  {Brinchmann}, {Charlot}, {Tremonti} \& {Brinkmann}}{{Kauffmann}
  et~al.}{2004}]{Kauffmann04}
{Kauffmann} G.,  {White} S.~D.~M.,  {Heckman} T.~M.,  {M{\'e}nard} B.,
  {Brinchmann} J.,  {Charlot} S.,  {Tremonti} C.,    {Brinkmann} J.,  2004,
  \mnras, 353, 713

\bibitem[\protect\citeauthoryear{{Khochfar} \& {Ostriker}}{{Khochfar} \&
  {Ostriker}}{2008}]{Khochfar08}
{Khochfar} S.,  {Ostriker} J.~P.,  2008, \apj, 680, 54

\bibitem[\protect\citeauthoryear{{Kravtsov}, {Berlind}, {Wechsler}, {Klypin},
  {Gottl{\"o}ber}, {Allgood} \& {Primack}}{{Kravtsov}
  et~al.}{2004}]{Kravtsov04}
{Kravtsov} A.~V.,  {Berlind} A.~A.,  {Wechsler} R.~H.,  {Klypin} A.~A.,
  {Gottl{\"o}ber} S.,  {Allgood} B.,    {Primack} J.~R.,  2004, \apj, 609, 35

\bibitem[\protect\citeauthoryear{{Li}, {Kauffmann}, {Jing}, {White},
  {B{\"o}rner} \& {Cheng}}{{Li} et~al.}{2006}]{Li06}
{Li} C.,  {Kauffmann} G.,  {Jing} Y.~P.,  {White} S.~D.~M.,  {B{\"o}rner} G.,
   {Cheng} F.~Z.,  2006, \mnras, 368, 21

\bibitem[\protect\citeauthoryear{{Li} \& {White}}{{Li} \& {White}}{2009}]{Li09}
{Li} C.,  {White} S.~D.~M.,  2009, \mnras, 398, 2177

\bibitem[\protect\citeauthoryear{{Macci{\`o}} \& {Fontanot}}{{Macci{\`o}} \&
  {Fontanot}}{2010}]{Maccio10}
{Macci{\`o}} A.~V.,  {Fontanot} F.,  2010, \mnras, 404, L16

\bibitem[\protect\citeauthoryear{{Mandelbaum}, {Seljak}, {Kauffmann}, {Hirata}
  \& {Brinkmann}}{{Mandelbaum} et~al.}{2006}]{Mandelbaum06}
{Mandelbaum} R.,  {Seljak} U.,  {Kauffmann} G.,  {Hirata} C.~M.,    {Brinkmann}
  J.,  2006, \mnras, 368, 715

\bibitem[\protect\citeauthoryear{{Marchesini}, {van Dokkum}, {F{\"o}rster
  Schreiber}, {Franx}, {Labb{\'e}} \& {Wuyts}}{{Marchesini}
  et~al.}{2009}]{Marchesini09}
{Marchesini} D.,  {van Dokkum} P.~G.,  {F{\"o}rster Schreiber} N.~M.,  {Franx}
  M.,  {Labb{\'e}} I.,    {Wuyts} S.,  2009, \apj, 701, 1765

\bibitem[\protect\citeauthoryear{{McCarthy}, {Frenk}, {Font}, {Lacey}, {Bower},
  {Mitchell}, {Balogh} \& {Theuns}}{{McCarthy} et~al.}{2008}]{McCarthy08}
{McCarthy} I.~G.,  {Frenk} C.~S.,  {Font} A.~S.,  {Lacey} C.~G.,  {Bower}
  R.~G.,  {Mitchell} N.~L.,  {Balogh} M.~L.,    {Theuns} T.,  2008, \mnras,
  383, 593

\bibitem[\protect\citeauthoryear{{Mihos} \& {Hernquist}}{{Mihos} \&
  {Hernquist}}{1994}]{Mihos94}
{Mihos} J.~C.,  {Hernquist} L.,  1994, \apjl, 431, L9

\bibitem[\protect\citeauthoryear{{Moster}, {Somerville}, {Maulbetsch}, {van den
  Bosch}, {Macci{\`o}}, {Naab} \& {Oser}}{{Moster} et~al.}{2010}]{Moster10}
{Moster} B.~P.,  {Somerville} R.~S.,  {Maulbetsch} C.,  {van den Bosch} F.~C.,
  {Macci{\`o}} A.~V.,  {Naab} T.,    {Oser} L.,  2010, \apj, 710, 903

\bibitem[\protect\citeauthoryear{{Neistein} \& {Weinmann}}{{Neistein} \&
  {Weinmann}}{2010}]{Neistein10}
{Neistein} E.,  {Weinmann} S.~M.,  2010, \mnras, 405, 2717 (NW10)

\bibitem[\protect\citeauthoryear{{Panter}, {Jimenez}, {Heavens} \&
  {Charlot}}{{Panter} et~al.}{2007}]{Panter07}
{Panter} B.,  {Jimenez} R.,  {Heavens} A.~F.,    {Charlot} S.,  2007, \mnras,
  378, 1550

\bibitem[\protect\citeauthoryear{{P{\'e}rez-Gonz{\'a}lez}, {Rieke}, {Villar},
  {Barro}, {Blaylock}, {Egami}, {Gallego}, {Gil de Paz}, {Pascual}, {Zamorano}
  \& {Donley}}{{P{\'e}rez-Gonz{\'a}lez} et~al.}{2008}]{PerzGonzalez08}
{P{\'e}rez-Gonz{\'a}lez} P.~G.,  {Rieke} G.~H.,  {Villar} V.,  {Barro} G.,
  {Blaylock} M.,  {Egami} E.,  {Gallego} J.,  {Gil de Paz} A.,  {Pascual} S.,
  {Zamorano} J.,    {Donley} J.~L.,  2008, \apj, 675, 234

\bibitem[\protect\citeauthoryear{{Sawala}, {Guo}, {Scannapieco}, {Jenkins} \&
  {White}}{{Sawala} et~al.}{2010}]{Sawala10}
{Sawala} T.,  {Guo} Q.,  {Scannapieco} C.,  {Jenkins} A.,    {White} S.~D.~M.,
  2010, ArXiv e-prints, 1003.0671

\bibitem[\protect\citeauthoryear{{Shankar}, {Lapi}, {Salucci}, {De Zotti} \&
  {Danese}}{{Shankar} et~al.}{2006}]{Shankar06}
{Shankar} F.,  {Lapi} A.,  {Salucci} P.,  {De Zotti} G.,    {Danese} L.,  2006,
  \apj, 643, 14

\bibitem[\protect\citeauthoryear{{Somerville}, {Primack} \&
  {Faber}}{{Somerville} et~al.}{2001}]{Somerville01}
{Somerville} R.~S.,  {Primack} J.~R.,    {Faber} S.~M.,  2001, \mnras, 320, 504

\bibitem[\protect\citeauthoryear{{Springel}, {White}, {Jenkins}, {Frenk},
  {Yoshida}, {Gao}, {Navarro}, {Thacker}, {Croton}, {Helly}, {Peacock}, {Cole},
  {Thomas}, {Couchman}, {Evrard}, {Colberg} \& {Pearce}}{{Springel}
  et~al.}{2005}]{Springel05}
{Springel} V.,  {White} S.~D.~M.,  {Jenkins} A.,  {Frenk} C.~S.,  {Yoshida} N.,
   {Gao} L.,  {Navarro} J.,  {Thacker} R.,  {Croton} D.,  {Helly} J.,
  {Peacock} J.~A.,  {Cole} S.,  {Thomas} P.,  {Couchman} H.,  {Evrard} A.,
  {Colberg} J.,    {Pearce} F.,  2005, \nat, 435, 629

\bibitem[\protect\citeauthoryear{{Springel}, {White}, {Tormen} \&
  {Kauffmann}}{{Springel} et~al.}{2001}]{Springel01}
{Springel} V.,  {White} S.~D.~M.,  {Tormen} G.,    {Kauffmann} G.,  2001,
  \mnras, 328, 726

\bibitem[\protect\citeauthoryear{{Tinker}, {Weinberg}, {Zheng} \&
  {Zehavi}}{{Tinker} et~al.}{2005}]{Tinker05}
{Tinker} J.~L.,  {Weinberg} D.~H.,  {Zheng} Z.,    {Zehavi} I.,  2005, \apj,
  631, 41

\bibitem[\protect\citeauthoryear{{Vale} \& {Ostriker}}{{Vale} \&
  {Ostriker}}{2004}]{Vale04}
{Vale} A.,  {Ostriker} J.~P.,  2004, \mnras, 353, 189

\bibitem[\protect\citeauthoryear{{van den Bosch}, {Aquino}, {Yang}, {Mo},
  {Pasquali}, {McIntosh}, {Weinmann} \& {Kang}}{{van den Bosch}
  et~al.}{2008}]{vdBosch08}
{van den Bosch} F.~C.,  {Aquino} D.,  {Yang} X.,  {Mo} H.~J.,  {Pasquali} A.,
  {McIntosh} D.~H.,  {Weinmann} S.~M.,    {Kang} X.,  2008, \mnras, 387, 79

\bibitem[\protect\citeauthoryear{{Wang}, {Li}, {Kauffmann} \& {De
  Lucia}}{{Wang} et~al.}{2006}]{Wang06}
{Wang} L.,  {Li} C.,  {Kauffmann} G.,    {De Lucia} G.,  2006, \mnras, 371, 537

\bibitem[\protect\citeauthoryear{{Wechsler}, {Zentner}, {Bullock}, {Kravtsov}
  \& {Allgood}}{{Wechsler} et~al.}{2006}]{Wechsler06}
{Wechsler} R.~H.,  {Zentner} A.~R.,  {Bullock} J.~S.,  {Kravtsov} A.~V.,
  {Allgood} B.,  2006, \apj, 652, 71

\bibitem[\protect\citeauthoryear{{Weinmann}, {Kauffmann}, {van den Bosch},
  {Pasquali}, {McIntosh}, {Mo}, {Yang} \& {Guo}}{{Weinmann}
  et~al.}{2009}]{Weinmann09}
{Weinmann} S.~M.,  {Kauffmann} G.,  {van den Bosch} F.~C.,  {Pasquali} A.,
  {McIntosh} D.~H.,  {Mo} H.,  {Yang} X.,    {Guo} Y.,  2009, \mnras, 394, 1213

\bibitem[\protect\citeauthoryear{{Weinmann}, {Kauffmann}, {von der Linden} \&
  {De Lucia}}{{Weinmann} et~al.}{2010}]{Weinmann10}
{Weinmann} S.~M.,  {Kauffmann} G.,  {von der Linden} A.,    {De Lucia} G.,
  2010, \mnras, 406, 2249

\bibitem[\protect\citeauthoryear{{Yang}, {Mo} \& {van den Bosch}}{{Yang}
  et~al.}{2009}]{Yang09}
{Yang} X.,  {Mo} H.~J.,    {van den Bosch} F.~C.,  2009, \apj, 695, 900

\bibitem[\protect\citeauthoryear{Zehavi et~al.,}{Zehavi
  et~al.}{2005}]{Zehavi05}
Zehavi I.,  et~al., 2005, \apj, 630, 1

\end{thebibliography}

\section*{Appendix}

\begin{figure*}
\centerline{\psfig{file=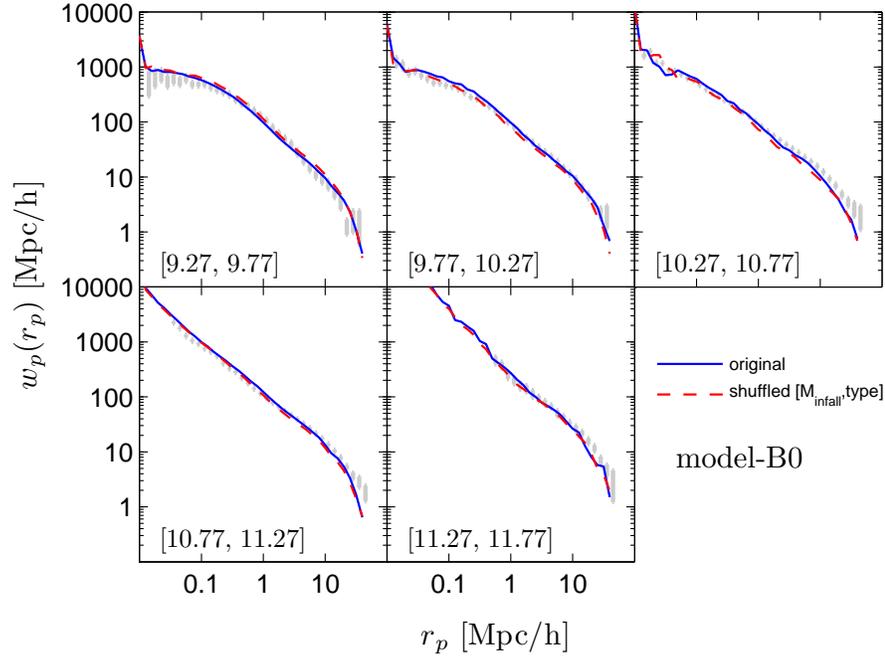,width=130mm,bbllx=30mm,bblly=80mm,bburx=188mm,bbury=200mm,clip=}}
\caption{The projected auto-correlation functions derived for model $B0$.
Each panel corresponds to galaxies with stellar masses as indicated, in units of Log$\msun$.
Solid lines show the results of the original model, dashed lines are
plotted using shuffling within [$\mfal$,type]. The observational
data are using SDSS DR7 with the same technique as in \citet{Li06}, and are shown as
error bars.}
\label{fig:cf_shuffled_B0}
\end{figure*}

\begin{figure*}
\centerline{\psfig{file=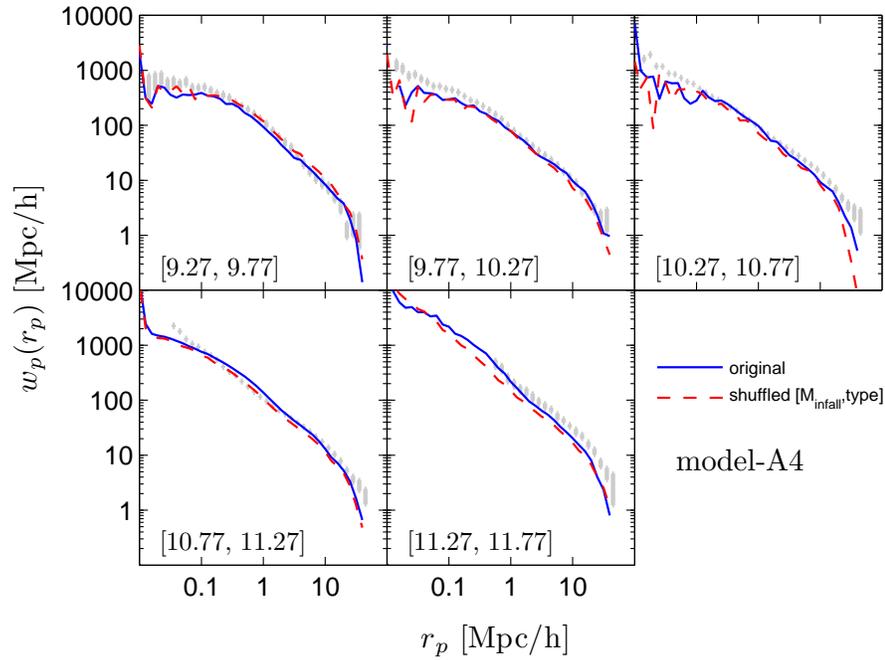,width=130mm,bbllx=30mm,bblly=80mm,bburx=188mm,bbury=200mm,clip=}}
\caption{Similar to Fig.~\ref{fig:cf_shuffled_B0}, but for model $A4$.}
\label{fig:cf_shuffled_A4}
\end{figure*}

\begin{figure*}
\centerline{\psfig{file=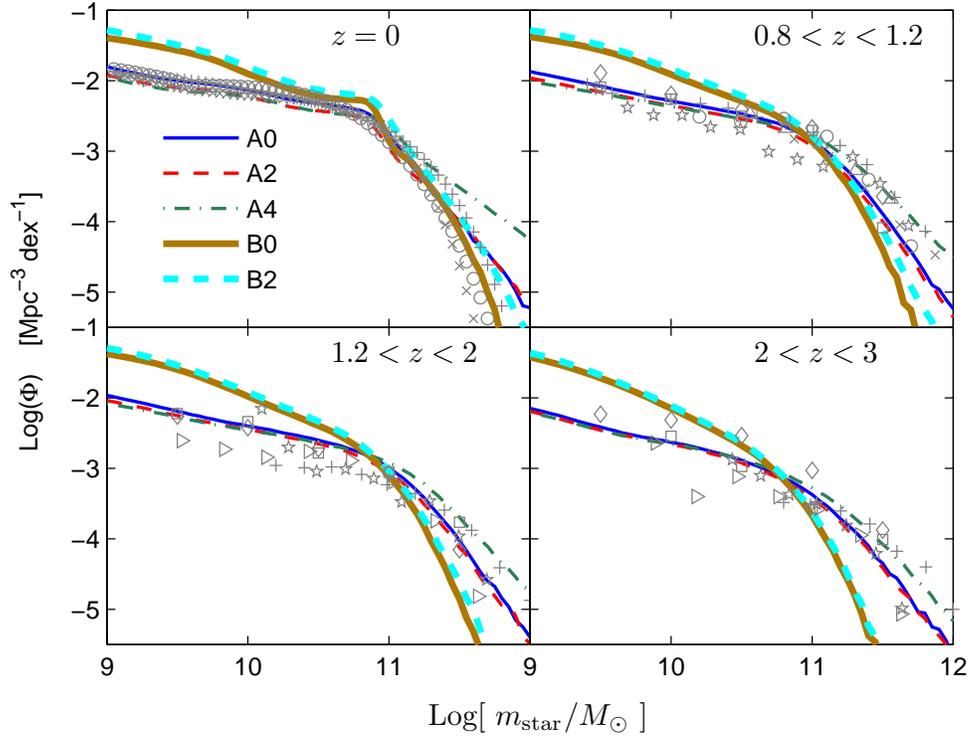,width=150mm,bbllx=30mm,bblly=80mm,bburx=188mm,bbury=200mm,clip=}}
\caption{Stellar mass functions of galaxies at various redshifts, for our different models as indicated in
the top left panel and as summarized in table \ref{tab:models}. The
observational data is plotted with gray symbols. At all redshifts
higher than zero we convolve the model stellar masses with a
Gaussian error distribution, with standard deviation of 0.25 dex. At
$\z=0$ we use observations by \citet[][circles]{Li09}, \citet[][crosses]{Baldry08},
\citet[][pluses]{Panter07}. Note that other observational studies predict a slightly
different function \citep[e.g.][]{Bernardi10}. At high-$\z$ we use the following
observations: \citet[][$\z=0.75-1$, circles]{Bundy06},
\citet[][$\z=0.8-1$, crosses]{Borch06}, \citet[][$\z=0.8-1$,
$\z=1.6-2$, $\z=2.5-3$, plus signs]{PerzGonzalez08},
\citet[][$\z=0.8-1$, $\z=1.6-2$, $\z=2-3$, stars]{Fontana06},
\citet[][$\z=0.8-1$, upward-pointing triangles]{Drory04},
\citet[][$\z=0.75-1.25$, $\z=1.75-2.25$, $z=2.25-3$, diamonds and
squares]{Drory05}, \citet[][$\z=1.3-2$, $\z=2-3$, right-pointing
triangles]{Marchesini09}. Model stellar mass functions are plotted
at $\z=0$, 1, 1.5, 2.5 according to the label on each panel. We
treat the specific IMF chosen in each measurement as part of the
observational `uncertainty' and do not convert them into
the same IMF.}
\label{fig:mass_funs}
\end{figure*}

In this Appendix we provide more CF plots for models $B0$ and $A4$ (Figs.~\ref{fig:cf_shuffled_B0},
\ref{fig:cf_shuffled_A4}). We also plot the stellar mass function for all the models used in this work
against the observational data in Fig.~\ref{fig:mass_funs}.

\label{lastpage}

\end{document}